\documentclass[conference]{IEEEtran}
\IEEEoverridecommandlockouts
\usepackage{cite}
\usepackage{amsmath,amssymb,amsfonts}
\usepackage{algorithmic}
\usepackage{algorithm}  
\usepackage{graphicx}
\usepackage{textcomp}
\usepackage{xcolor}
\usepackage{float} 
\usepackage{bm}
\usepackage{multirow}
\usepackage{array}
\usepackage{booktabs}
\usepackage{multirow}
\usepackage{fancyhdr} % 添加页眉页脚
\usepackage{color}
\usepackage{tabu}
\usepackage{nomencl}
\makenomenclature
\usepackage{etoolbox}
\usepackage{graphicx}
\usepackage{subfigure}
\usepackage[cal=boondoxo]{mathalfa}
\newcolumntype{I}{!{\vrule width 1.3pt}}
\newlength\savedwidth

\newlength\savewidth

\def\BibTeX{{\rm B\kern-.05em{\sc i\kern-.025em b}\kern-.08em
    T\kern-.1667em\lower.7ex\hbox{E}\kern-.125emX}}

\makeatletter
\patchcmd{\@maketitle}
  {\addvspace{0.5\baselineskip}\egroup}
  {\addvspace{0\baselineskip}\egroup}
  {}
  {}
\makeatother

\begin{document}

\title{Multi-agent Bayesian Deep Reinforcement Learning for Microgrid Energy Management under Communication Failures\\}

\author{\IEEEauthorblockN{Hao Zhou, Atakan Aral, \IEEEmembership{Member, IEEE}, Ivona Brandic, \IEEEmembership{Member, IEEE}, and Melike Erol-Kantarci, \IEEEmembership{Senior Member, IEEE}}

\thanks{ This work is funded by Natural Sciences and Engineering Research Council of Canada (NSERC), Collaborative Research and Training Experience Program (CREATE) under Grant 497981, by the CHIST-ERA grant CHIST-ERA-19-CES-005, and by the Austrian Science Fund (FWF): projects Y904-N31 (RUCON) and I5201-N (SWAIN).

H. Zhou and M. Erol-Kantarci are with the School of Electrical Engineering and Computer Science, University of Ottawa, Ottawa, ON K1N 6N5, Canada. (email:\{hzhou098, melike.erolkantarci\}@uottawa.ca) 

A. Aral is with the Faculty of Computer Science, University of Vienna, Vienna, A-1090, Austria. (e-mail: atakan.aral@univie.ac.at)

I. Brandic is with the Faculty of Informatics, Vienna University of Technology, Vienna, A-1040, Austria. (e-mail: ivona.brandic@tuwien.ac.at) 

Copyright (c) 20xx IEEE. Personal use of this material is permitted. However, permission to use this material for any other purposes must be obtained from the IEEE by sending a request to pubs-permissions@ieee.org.}}

\maketitle
\IEEEpubidadjcol
\pagestyle{empty}  % no page number for the second and the later pages
\thispagestyle{empty} % no page number for the first page

\maketitle
 \thispagestyle{fancy} % 
      \lhead{} % 页眉左，需要东西的话就在{}内添加
      \chead{} % 页眉中
      \rhead{} % 页眉右
      \lfoot{} % 页眉左
      \cfoot{} % 页眉中
      \rfoot{} %页眉右，\thepage 表示当前页码
      \renewcommand{\headrulewidth}{0pt} %改为0pt即可去掉页眉下面的横线
      \renewcommand{\footrulewidth}{0pt} %改为0pt即可去掉页脚上面的横线
\pagestyle{fancy}
\fancyhead[C]{Accepted by IEEE Internet of Things Journal for future publication, copyright belongs to @ 20xx IEEE. }

\begin{abstract}
Microgrids (MGs) are important players for the future transactive energy systems where a number of intelligent Internet of Things (IoT) devices interact for energy management in the smart grid. Although there have been many works on MG energy management, most studies assume a perfect communication environment, where communication failures are not considered. In this paper, we consider the MG as a multi-agent environment with IoT devices in which AI agents exchange information with their peers for collaboration. However, the collaboration information may be lost due to communication failures or packet loss. Such events may affect the operation of the whole MG. To this end, we propose a multi-agent Bayesian deep reinforcement learning (BA-DRL) method for MG energy management under communication failures. We first define a multi-agent partially observable Markov decision process (MA-POMDP) to describe agents under communication failures, in which each agent can update its beliefs on the actions of its peers. Then, we apply a double deep Q-learning (DDQN) architecture for Q-value estimation in BA-DRL, and propose a belief-based correlated equilibrium for the joint-action selection of multi-agent BA-DRL. Finally, the simulation results show that BA-DRL is robust to both power supply uncertainty and communication failure uncertainty. BA-DRL has 4.1\% and 10.3\% higher reward than Nash Deep Q-learning (Nash-DQN) and alternating direction method of multipliers (ADMM) respectively under 1\% communication failure probability.

\end{abstract}


\section{Introduction}
\label{s1}
A microgrid (MG) is a low-voltage, small scale power system, which usually contains one or more generators, user loads, and the energy storage system (ESS). In modern grids, these entities are interconnected and form an IoT environment where power loads, storage and generation communicate and collaborate for improved energy management. A number of centralized algorithms have been proposed for MG energy management, including the gradient search, interior point algorithm and lambda iteration approach \cite{b1}. However, these centralized methods raise concerns about reliability and security issues, such as the failure of central controller, which may result in shutdown of the whole MG. In addition, resilience, privacy protection, high computation burden, and high bandwidth requirements are all further challenges for a centralized MG architecture. 

Similar to the decentralization trend in many areas of resource orchestration, decentralized control is regarded as the future of MG energy management due to high flexibility and reliability, as well as low computation burden \cite{b2}. The decentralized MG architecture can be modeled as a multi-agent system, which includes a generator (e.g., photovoltaic power (PV)  agent), a demand side management (DSM) agent, an ESS agent, and so on. In a multi-agent system, multiple intelligent agents work collaboratively, and each agent optimizes its own objective in a parallel or sequential manner. Having physically separate units with different objectives and reasonable autonomy makes the MG an ideal example of a multi-agent system. 

In recent years, multi-agent based MG energy management techniques are widely studied with various methodologies, including the alternating direction method of multipliers (ADMM)\cite{b3}, event-trigger\cite{b4}, consensus theory \cite{b5}, and game theory \cite{b5-1}. 
On one hand, model-based methods generally require dedicated optimization models. For example, to deploy the convex optimization, the convexity should be first checked, and the non-convex problems need to be converted to convex optimizations, which may lead to great complexity \cite{b5-11}. On the other hand, the uncertainty of renewable energy resources, heterogeneous cost definitions of agents, different energy trading strategies are all potential challenges for model-based solutions, since they will greatly increase the difficulty of building a optimization model \cite{b5-12}. The limitations of traditional methods and their challenges motivate us to find more ideal solutions for the MG energy management. Recently, data-driven approaches, such as reinforcement learning, have emerged as a model-free alternative where machine learning algorithms are used for the management of MGs \cite{b5-13}. 
In particular, multi-agent reinforcement learning (MARL) has become a promising method for MG energy management and control\cite{b6}. Various reinforcement learning algorithms have been used for MG energy management in the literature, e.g., Bayesian reinforcement learning \cite{b5-2}, Nash Q-learning\cite{b7}, fuzzy Q-learning\cite{b8}, cooperative reinforcement learning\cite{b9} and deep Q-learning (DQN) \cite{b10}.

It is worth noting that communication between agents is the key for a multi-agent system in both model-based and model-free methods. By exchanging collaboration information, agents can choose optimal joint-action and maintain a satisfactory overall performance. On the contrary, missing collaboration information may lead to suboptimal decisions by agents, and in some cases, may even impact the stability of the whole MG \cite{b11, b11-1}. However, most existing research assumes perfect communication without information loss \cite{b3,b4,b5,b6,b7,b8,b9,b10}. 
As bandwidth, latency and reliability requirements of different smart grid applications vary widely, packet loss may happen while agents exchange the collaboration information \cite{b10-1}. The communication system can be prone to errors and signal loss due to harsh smart grid environment which contribute to high packet loss rate \cite{b10-2}. Consequently, it is critical to involve the packet loss in the MG energy management to guarantee the MG performance and reduce potential losses.

To the best of our knowledge, this is the first study to investigate the MG energy management under communication failures. Furthermore, the power supply uncertainty, especially the uncertain PV power generation, is generally regarded as an important challenge for MG operation \cite{b5-2,b6,b7}. The fluctuation of PV power increases the complexity of MG operation, and the proposed solution is expected to handle both power supply and communication system uncertainty. 

In this work, we propose a multi-agent Bayesian deep reinforcement learning (BA-DRL) algorithm for MG energy management. The proposed algorithm is designed to be robust under both power supply uncertainty and communication failures uncertainty between IoT-enabled loads, storage and generation in MG. We assume each agent employs BA-DRL independently and they share Q-values for collaboration, but the shared Q-values may be lost due to failures in the underlying communication system. These failures can be system failures or simply packet loss or delayed/obsolete packets. To address this problem, we include a novel Bayesian belief update method in our BA-DRL scheme, which estimates the actions of other agents by updating beliefs. Considering the large state-action space, a double deep Q-learning architecture is applied for Q-values estimation, and a novel belief-based correlated equilibrium is proposed for collaborative action selection of agents. The BA-DRL is compared to both model-free Nash deep Q-learning (Nash-DQN) and model-based ADMM methods, and the results show that BA-DRL maintains superior performance than baseline algorithms. 

The main contributions of this work are: 

(1) A multi-agent partially observable Markov decision process (MA-POMDP) based scheme is proposed to describe agents with communication failures. Based on MA-PODMP scheme, we propose a Bayesian belief update method, where each agent can update its beliefs on the actions of peer agents by Bayesian rules, and no direct communication is needed for the belief updating.  

(2) We propose the BA-DRL algorithm to handle the communication failures between agents in an MG. In BA-DRL, a double deep Q-learning (DDQN) architecture is applied for Q-value approximation to prevent over-estimation. We also propose a novel action selection method called belief-based correlated equilibrium. It utilizes beliefs to select optimal joint-action under communication failures, which is implemented in a decentralized manner.

(3) The proposed BA-DRL solution is compared with the state-of-the-art Nash-DQN and ADMM methods. The simulation results show that the proposed BA-DRL outperforms baseline algorithms. It achieves 4.1\% and 10.3\% higher reward compared with Nash-DQN and ADMM algorithms under 1\% communication failure probability, respectively.

The rest of this paper is organized as follows. Section \ref{s2} summarizes the related work. Section \ref{s3} presents MG system architecture, MG energy trading model, and problem formulation. Section \ref{s4} defines the MA-POMDP, and Section \ref{s5} introduces the BA-DRL and baseline algorithms. We show the results in Section \ref{s6}, and Section \ref{s7} concludes the paper. 

\section{Related Work}
\label{s2}
Recently there have been a significant number of studies on MARL based MG energy management \cite{b7, b8, b9,b10,b5-2}. For instance, power loss in energy trading is minimized in \cite{b5-2}, where the coalition information is transferred between MGs. A decentralized MG energy management method is proposed in \cite{b7}, where Nash equilibrium is used for coordination by sharing utility values. Furthermore, fuzzy Q-learning is applied in \cite{b8} for MG energy management. A diffusion strategy is used in \cite{b9} for cooperative reinforcement learning where the agent only needs to communicate with its neighbors. Considering the huge state-action space of MG, DQN is applied for energy trading in \cite{b10}, and the utility of MGs is balanced by Nash equilibrium. In most of the prior works, communication has been a key element as the agents collaborate by sharing critical information, e.g., coalition information \cite{b5-2}, utility values\cite{b7,b10}, state variables\cite{b8,b9} to make decisions. 

The underlying communication system is not always reliable in practice, and IoT devices in the MG may experience system failures, packet loss and delayed packets. Although MG energy management has been investigated by various models and algorithms, the communication failure problem has not been considered before.
Note that, in \cite{b11}, the effect of communication delay on MG economic dispatch is investigated, and it presents that communication delay will lead to a system performance fluctuation. \cite{b11-1} proposed a novel consensus-based economic dispatch algorithm to handle communication delay in MGs, where the communication delay results in a higher cost for MGs. The work of \cite{b11} and \cite{b11-1} show that it is necessary to consider the effect of communication systems in the MG energy management. However, to the best of our knowledge, the impacts of communication failure on MG energy management has not been investigated before. We have proposed a MARL scheme for MG energy management in \cite{b11-2}, and a multi-agent DQN method in \cite{b11-3}. But communication related problems were still not studied in those works. Yet, handling such uncertainty brings substantial performance improvements as demonstrated in systems other than MGs such as electric vehicles \cite{b11-32}.

In this paper, different than existing works, we investigate the MG energy management performance under both power supply uncertainty and communication failures uncertainty. In BA-DRL, our MG agents collaboratively learn their actions, and they further aim to make a good response when some agents are isolated by communication failures. Meanwhile, BA-DRL should provide a satisfying performance under power supply uncertainty.

\section{MG System Architecture and Energy Model}
\label{s3}
\subsection{System Architecture}

\begin{figure}[!t]
\setlength{\abovecaptionskip}{0pt} 
\centering
\includegraphics[width=8.5cm,height=4.8cm]{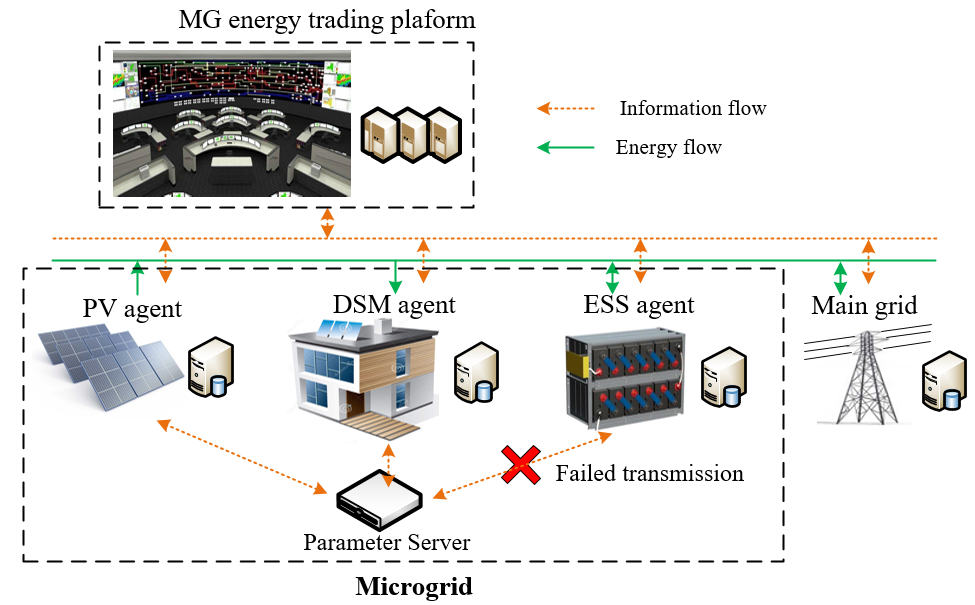}
\caption{Microgrid system architecture.}
\label{fig1}
\end{figure}

The MG system architecture is shown in Fig.\ref{fig1}. We consider three types of agents, namely PV, DSM, and ESS agents. A PV agent may control multiple solar panels, ESS agent can control multiple storage devices, and DSM agent controls all controllable loads. Agents exchange information with the energy trading platform to sell or buy energy, and maximize their own profit. Meanwhile, to maintain the overall benefit of the whole MG, it is reasonable to assume that agents can exchange collaboration information with each other by a parameter server. 
In each time slot,  we assume the the parameter server can forward the message of one agent to all other agents by a wireless network because of its low cost and high flexibility. This way all agents share the collaboration information with each other, which will be used for decision making of each agent \cite{b11-33}. However, communication failures may occur during information exchange, especially over the wireless network, and some agents may be temporarily isolated. 
Due to the missing collaboration information, the isolated agent is very likely to take arbitrary actions and the benefit of other agents may be harmed. As a result, it is critical to consider the communication failures between agents in MG energy management. 

\subsection{MG Agents Model}
The DSM agent is assumed to control the demand of multiple deferrable devices such as water heater or dishwasher. The devices connect to the DSM controller via an IoT technology as in \cite{b11-4}. These devices can change their operation time to reduce costs, and the comfort of end users will not be affected. We assume an average power consumption, and the total power demand of DSM agents with $D$ sets of devices is:
\begin{equation}
\setlength\abovedisplayskip{3pt}
\setlength\belowdisplayskip{3pt}
 \label{eu_eqn}
P^{dsm}_{t} =\sum_{g=1}^{D}P_{g}G_{t,g}
\end{equation}
where $D$ is the total number of DSM devices sets, $P_{g}$ is the average power consumption of devices in set $g$, $G_{t,g}$ is a binary value to represent the on/off status. 
Considering the end users' comfort level, these devices must be serviced before a certain time limit. The waiting time of devices is described as:
\begin{equation}
\setlength\abovedisplayskip{3pt}
\setlength\belowdisplayskip{3pt}
\vec W_{t}=[W_{t,1},W_{t,2},\dots,W_{t,g},\dots, W_{t,D}],
\end{equation}
where $W_{t,g}$ is the waiting time of device $g$ at time $t$. Here we assume the device $g$ should be serviced between the operation time limit $[t_{start},t_{end}]$, which means $W_{t,g}\leq t_{end}-t_{start}$, and $W_{t,g}$ can be easily calculated by $W_{t,g}=t-t_{start}$, in which $t$ is current time slot. We set $W_{t,g}$ to 0 when $t<t_{start}$ or $t>t_{end}$.
Meanwhile, crucial devices, e.g., lighting and cooking devices, which are unable to change their operation time, are not involved in this work since the agent has no control over them.

For the PV agent, we assume the PV power can be predicted with an error term, which describes the uncertainty of the PV power and the possible prediction inaccuracy\cite{b11-5}.

The power of ESS agent is: 
\begin{equation}
\setlength\abovedisplayskip{3pt}
\setlength\belowdisplayskip{1pt}
\label{eu_eqn}
P^{ess}_{t} =P^{char}q_{t}
\end{equation}
where $P^{char}$ is the fixed charging power, and $q_{t}$ is equal to 1, -1 and 0 when discharge, charge, and unchanged, respectively. Note that we assume a centralized MG-level ESS, instead of one ESS for each house in MG \cite{b11-6}.  

The state of charge (SOC) of ESS is updated according to: 
\begin{equation}
\setlength\abovedisplayskip{3pt}
\label{eu_eqn}
SOC_{t+1} =SOC_{t}-\frac{P^{char}}{C^{ess}}q_{t}
\setlength\belowdisplayskip{3pt}
\end{equation}
where $C^{ess}$ is the ESS capacity.

\subsection{Bidding based MG Energy Trading}
\begin{figure}[!t]
\setlength{\abovecaptionskip}{3pt} 
\centering
\includegraphics[width=8.5cm,height=4.5cm]{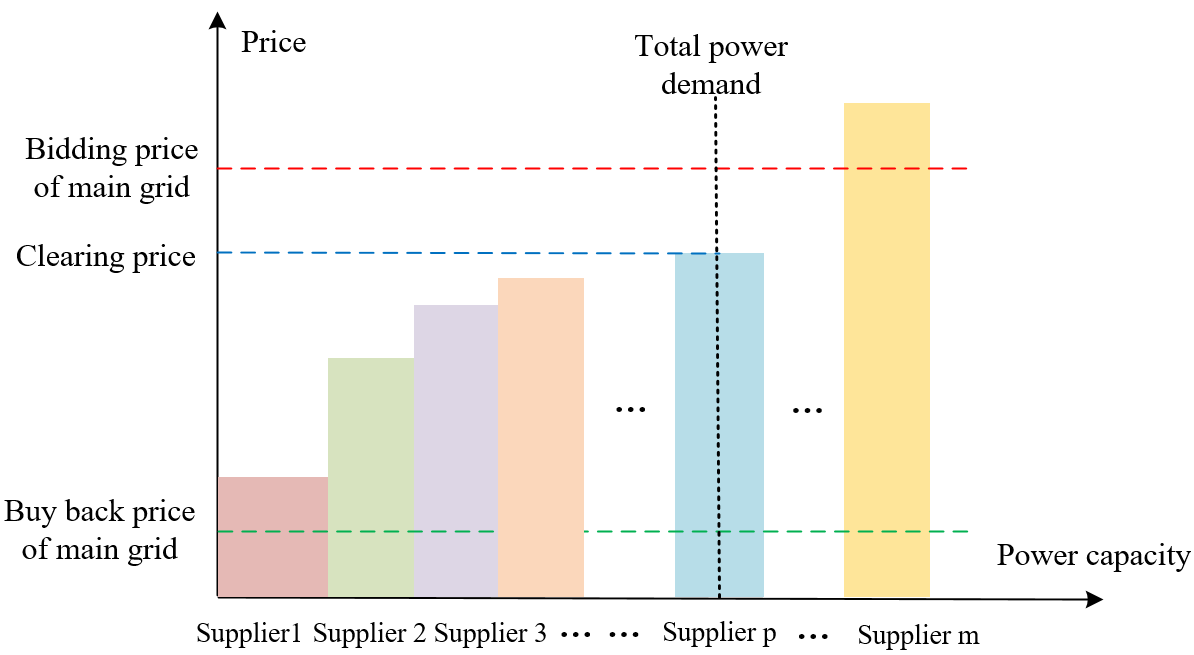}
\caption{Bidding based MG energy trading.}
\label{fig1-2}
\end{figure}

In this section, we introduce the MG energy trading model, where power suppliers and consumers are involved\cite{b11-7}. In the proposed MG, PV agent is a power supplier, and DSM agent is a power consumer, while ESS agent can be a supplier when discharging, or consumer when charging. The main grid can also participate the energy trading as a supplier or consumer for the energy balance of MG. Firstly, every supplier submits its power supply capacity $P^{sup}$ and bidding price $p^{sup}$. All consumers submit their power demand $P^{dem}$, and total power demand is calculated accordingly. Next, as shown in Fig.\ref{fig1-2}, the suppliers will be ranked from lowest to highest bidding price. Then, the bidding price at the intersection of total power demand and total supply is determined as the clearing price for this market, which indicates total power supply is equal with power demand with this price.

Moreover, if $\sum P^{sup}<\sum P^{dem}$, the main grid participates in the market as a supplier, and the bidding price $p^{grid}$, which is offered by main grid, will be the clearing price. On the contrary, if $\sum P^{sup}>\sum P^{dem}$, some suppliers need to sell the surplus electricity to the main grid, which is normally at a lower price. The bidding price that is higher than the $p^{grid}$ will be unacceptable, because consumers always desire a lower energy price. Similarly, the suppliers are unlikely to offer a price that is lower than the price of selling energy directly to the main grid. In this market, suppliers with a lower bidding price are more likely to be accepted, which will benefit the MG consumers\cite{b12}. It is worth noting that the bidding information is private for each agent, and no agent is able to manipulate the market.

\subsection{Problem Formulation}
Considering the DSM agent is always a consumer, it needs to minimize its cost:
\begin{small}
\begin{equation}
\setlength\abovedisplayskip{3pt}
\setlength\belowdisplayskip{3pt}
\label{eq4}
min(\sum^{T^{op}}_{t=1}\sum_{g=1}^{D}P_{g}G_{t,g}p^{c}_{t})
\end{equation}
\end{small}
where $T^{op}$ is the optimization horizon, $p^{c}_{t}$ is the clearing price.

The PV agent, in turn, aims to maximize the utility:
\begin{equation}
\setlength\abovedisplayskip{3pt}
\setlength\belowdisplayskip{3pt}
\label{eq5}
\resizebox{0.85\hsize}{!}{$max(\sum^{T^{op}}_{t=1}({P}^{pv}_{t}p^{c}_{t}-(\beta({P}^{pv}_{t})^2+\zeta{P}^{pv}_{t}+\Phi)))$}
\end{equation}
where $P^{pv}_{t}$ is the PV power. The widely used quadratic function is adopted to present the generation cost of PV, and $\beta, \zeta, \Phi $ are cost coefficients\cite{b12-1}.

Finally, the ESS agent maximizes its utility using:
\begin{equation}
\setlength\abovedisplayskip{3pt}
\setlength\belowdisplayskip{3pt}
\label{eq6}
max(\sum^{T^{op}}_{t=1}P^{ess}_{t}p^{c}_{t})
\end{equation}

Equation (\ref{eq4}), (\ref{eq5}) and (\ref{eq6}) define the objectives of DSM, PV and ESS agents, respectively. The DSM agent controls the on/off status of its devices to minimize its energy cost as a consumer. PV agent intends to maximize its profit by submitting an appropriate bidding price as an energy supplier. ESS agent buys energy when the energy price is lower, and selling energy when the price is higher to maximize the total profit. Equation (\ref{eq4}) to (\ref{eq6}) show that each agent can make decisions autonomously to maximize its profit or reduce the cost, then all three objectives are simultaneously optimized.
Meanwhile, the problem is optimized under following constraints:
\begin{equation}
\setlength\abovedisplayskip{3pt}
 \label{10}
\resizebox{0.6\hsize}{!}{$P^{pv}_{t}+P^{grid}_{t}+P^{ess}_{t}=P^{dsm}_{t}$}
\setlength\belowdisplayskip{3pt}
\end{equation}
\begin{equation}
\setlength\abovedisplayskip{3pt}
\label{11}
G_{t,g}\leq W_{t,g}
\setlength\belowdisplayskip{3pt}
\end{equation}
\begin{equation}
\setlength\abovedisplayskip{3pt}
\label{11-1}
W_{t,g}\leq W^{max}_{g}
\setlength\belowdisplayskip{3pt}
\end{equation}
\begin{equation}
\setlength\abovedisplayskip{3pt}
\setlength\belowdisplayskip{3pt} \label{12}
SOC_{min}\leq SOC_{t} \leq SOC_{max}
\end{equation}

Equation (\ref{10}) is the energy balance constraint; (\ref{11}) is the DSM constraint, which means only devices that have not been serviced can be turned on; (\ref{11-1}) means the waiting time cannot exceed the maximum limit; (\ref{12}) is the SOC lower and upper bound constraint.

\section{Multi-agent POMDP}
\label{s4}
In this section, we define the MA-POMDP and introduce a Bayesian belief update method. In MA-POMDP, each agent maintains beliefs on other agents. When some agents are isolated by communication failures, other agents can make reasonable estimations about the actions of isolated agents.

\begin{figure}[!t]
\setlength{\abovecaptionskip}{0pt} 
\centering
\subfigure[MDP architecture]{
\includegraphics[width=7.9cm,height=1.5cm]{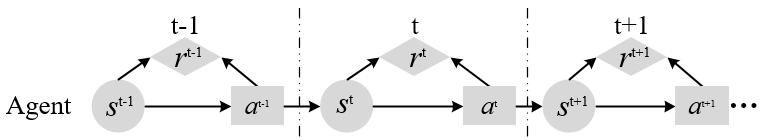}
}
\subfigure[POMDP architecture]{
\includegraphics[width=7.9cm,height=1.4cm]{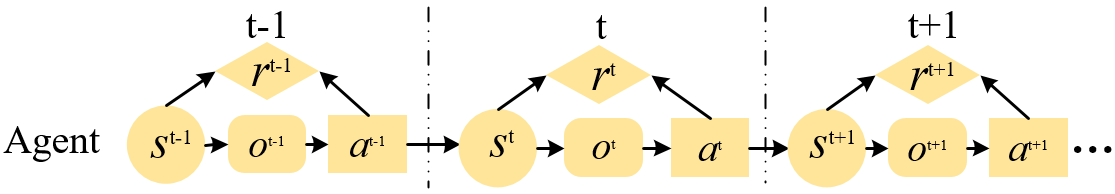}
}
\subfigure[Multi-agent POMDP architecture.]{
\includegraphics[width=8cm,height=7.4cm]{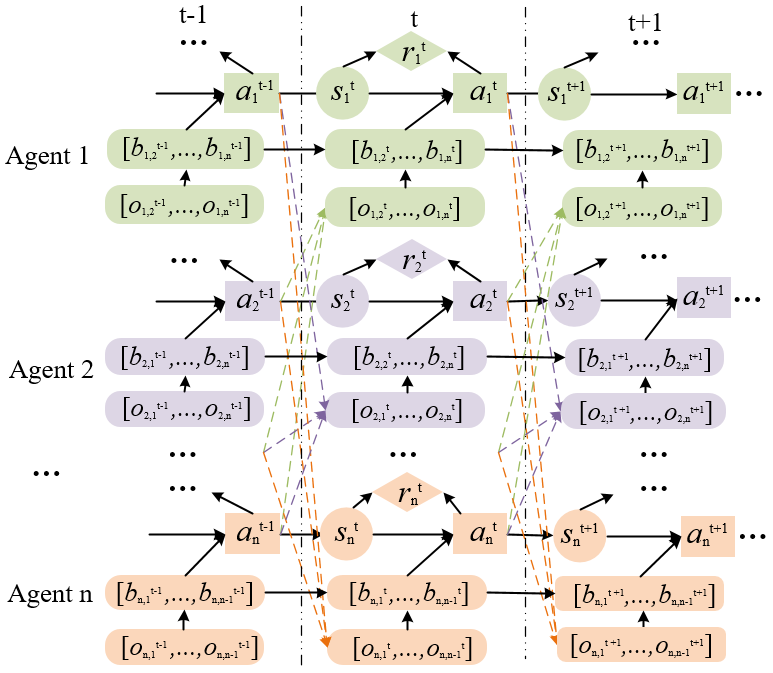}
}
\setlength{\abovecaptionskip}{0pt} 
\caption{MDP, POMDP and multi-agent POMDP comparison}
\label{fig1-3}
\end{figure}

\subsection{MA-POMDP Definition}
\label{s4-1}
As shown in Fig.\ref{fig1-3}(a), in the conventional MDP, an agent takes actions $a^{t}$ based on the current state $s^{t}$, then receives a reward $r^{t}$ and arrives to a new state $s^{t+1}$. However, in some cases, the agent cannot directly observe the underlying state $s^{t}$, which is common in practice. POMDP is proposed to describe cases in which the state is partially observable \cite{b13}. The POMDP is represented with $<S,A,T,R,\Omega,O,b >$, in which $S$, $A$, $T$ and $R$ represent the set of states, set of actions, transition probability, and reward function, respectively\cite{b14}. $\Omega$ is the set of observations $o$, $O$ is the observation function, and $b(s^{t})$ is the belief of state $s^{t}$. In POMDP, an agent is not sure about its state $s^{t}$, but the observation $o^{t}$ is always available when it comes to a new state. The observation function $O(s^{t}|o^{t})$ maps the observation result $o^{t}$ to state $s^{t}$. Fig.\ref{fig1-3}(b) shows the architecture of POMDP. It is worth noting that agents can only choose actions by observations $o^{t}$ in POMDP, instead of $s^{t}$ in MDP.

Although POMDP is defined for solving the state uncertainty, an agent in a multi-agent system is more likely to be uncertain about the actions of other agents instead of its own state. These unknown actions will directly affect system environment. The proposed MA-POMDP is shown in Fig.\ref{fig1-3}(c), where each agent has its own state, action, reward, observation and belief. Based on current state and beliefs, agent selects its own actions, and gets the corresponding reward, then it moves to the next state. Meanwhile, each agent will make observations about actions of its peers, which is represented by the dashed lines from the action of other agents to observations of one specific agent.

For agent $\lambda$, the observation set is denoted as 
$ \Vec{o}^{t}_{-\lambda}=(o^{t}_{\lambda,1},o^{t}_{\lambda,2},...,o^{t}_{\lambda,\lambda-1},o^{t}_{\lambda,\lambda+1},...,o^{t}_{\lambda,n})$. Here, $o^{t}_{\lambda,\varphi}$ means the observation made by the agent $\lambda$ for the action of agent $\varphi$ at time $t$ $(\varphi\neq\lambda)$. Indeed, one agent is unlikely to fully share its own action selection with other agents, especially under communication failure. However, the observations do not rely on actual communications, they are generated according to the accumulated experience of agents, which means no direct communication is needed \cite{b14-2}. Given this characteristic of observation function, we can utilize the MA-POMDP as a solution for communication failure between agents. Specifically, we assume that each intelligent agent can maintain a belief set $ \Vec{b}^{t}_{-\lambda}=(b^{t}_{\lambda,1},b^{t}_{\lambda,2},...,b^{t}_{\lambda,\lambda-1},b^{t}_{\lambda,\lambda+1},...,b^{t}_{\lambda,n})$ about actions of other $n-1$ agents.  $b^{t}_{\lambda,\varphi}$ means the belief of agent $\lambda$ for the action of agent $\varphi$ at time $t$ $(\varphi\neq\lambda)$. Based on observations at time $t$, the belief can be updated from $t$ to $t+1$, as described in following Section \ref{s4-2}. Noting that the proposed MA-POMDP is different with existing decentralized POMDP \cite{b14-2}. In decentralized POMDP, all agents work as a team to maximize the global reward, but each agent has its own reward in our proposed MA-POMDP.

Based on proposed MA-POMDP scheme, we define the state, action, reward, and observation function of MG agents:  
\begin{itemize}
 \item For the DSM agent, the state is defined by current time and device waiting time $\{t,\vec W\}$, and the action is the on/off status of DSM devices. The reward function is given by the objective of DSM agent to minimize energy cost, which is indicated by equation (5). 
\item For PV agent, the state is defined by current time and power $\{t, P_{t}^{PV}\}$, and the action is bidding price. The reward function is defined by equation (6) in the problem formulation to maximize its profit. 
\item For ESS agent, the state is defined by time and SOC $\{t,SOC\}$, and the action includes ESS power and bidding price. The reward function is defined as equation (7) to maximize its utility. 
\end{itemize}

\begin{figure}[!t]
\setlength{\abovecaptionskip}{0pt} 
\centering
\includegraphics[width=9cm,height=5cm]{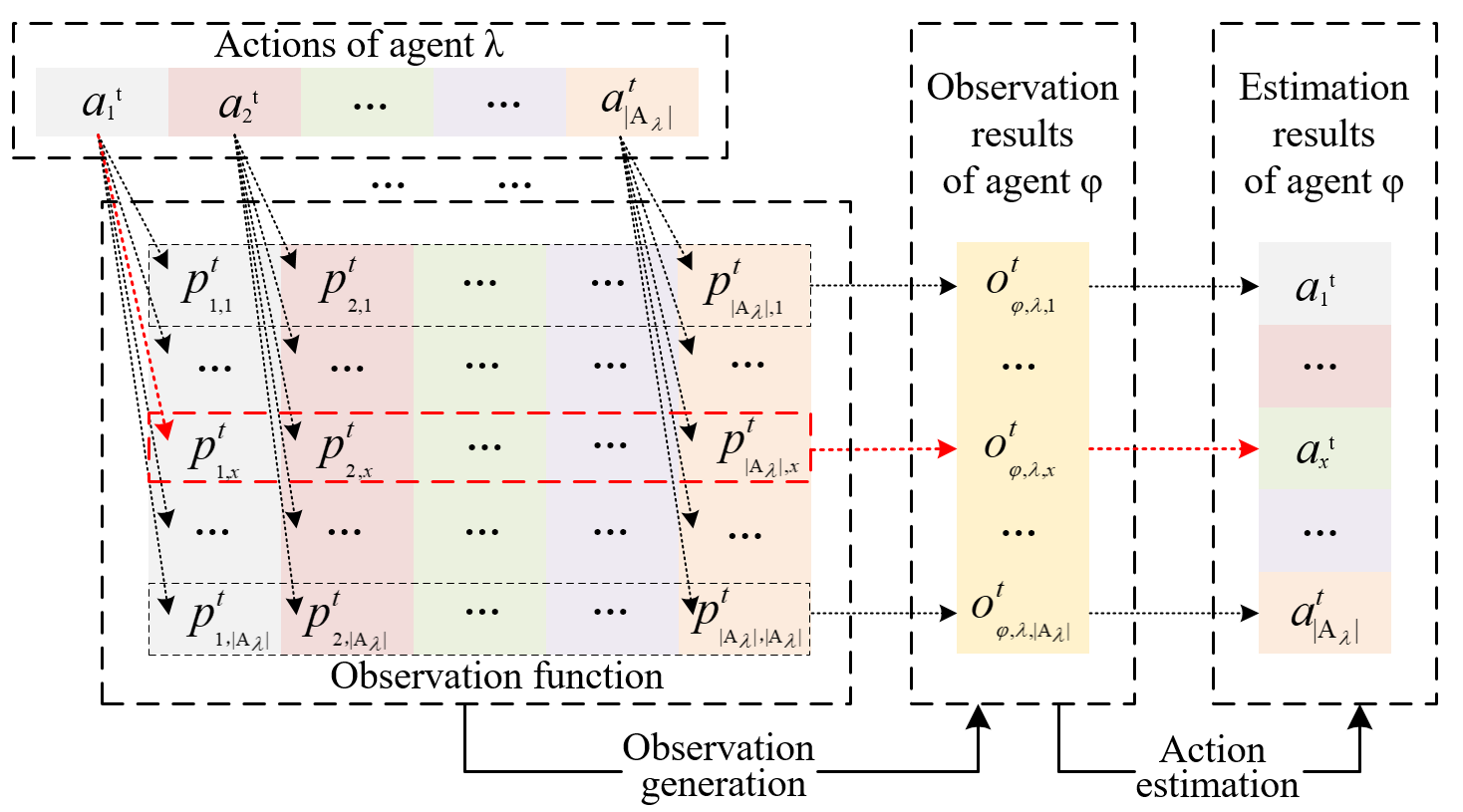}
\caption{Observation function definition.}
\label{fig1-11}
\end{figure}

The proposed observation function is presented as Fig.\ref{fig1-11}, which can be divided into two parts: observation generation and action estimation. Firstly, in the observation generation phase, when agent $\lambda$ takes action $a^{t}_{1}$, an intermediate observation $\mathcal{o}^{t}_{1,x}$ will be generated with a probability $p^{t}_{1,x}$ ($1\leq x \leq |A_{\lambda}|$, $\sum^{|A_{\lambda}|}_{x=1} p^{t}_{1,x}=1 $), in which the $\mathcal{o}^{t}_{1,x}$ means the $x^{th}$ intermediate observation of action $a^{t}_{1}$ (shown by the red dashed arrow in the left of Fig.\ref{fig1-11}). Then the $\mathcal{o}^{t}_{1,x}$ will be mapped to ${o}^{t}_{\varphi,\lambda,x}$, which is the final observation results received by agent $\varphi$. Secondly, in the action estimation phase, the observation results received by agent $\varphi$ will be mapped to the estimated actions of agent $\lambda$. As shown by Fig.\ref{fig1-11}, the relationship between observation and estimation results is defined by ${o}^{t}_{\varphi,\lambda,x} \rightarrow a^{t}_{x}$, which means receiving ${o}^{t}_{\varphi,\lambda,x}$ will lead to estimated action $a^{t}_{x}$.
However, ${o}^{t}_{1,\lambda,x}$ can also be mapped by other intermediate observations like $p^{t}_{2,x}$ or $p^{t}_{|A_{\lambda}|,x}$. It denotes that the real action of agent $\lambda$ is never known by agent $\varphi$, and observation error is inevitable.  For example, when DSM agent takes an action $a_{dsm}^{t}$, PV and ESS agents will receive observations for the on/off status of DSM agent, which is an estimated action of DSM agent. Although the real action $a_{dsm}^{t}$ is never known by PV and ESS agents, the long-term observation can still reflect the action selection routine of DSM agent, which can be utilized to update the beliefs. Note that the observation functions can be defined in various ways and here we define our function according to the MG environment. 

\subsection{Bayesian Belief Update Method}
\label{s4-2}
In this section, we introduce how to use observations to update beliefs by Bayesian rule. Let the agent with communication failures be called as problematic agent (PA), and other agents are normal agents (NAs). Note that the proposed method can be applied for the multiple PA case without loss of generality. Here PA is isolated due to failed transmissions (which NAs can infer by a missing periodic update), and it will ignore other agents when choosing its own actions. Meanwhile, considering all agents are independent, NAs are unable to know the state and actions of the PA, as well. However, NAs could make some observations about the actions of PA, and map the observation results to actions, which is a basic assumption in POMDP problems\cite{b13}. 

The historical experience from former $j-1$ episodes are used to form the beliefs of actions in $j^{th}$ episode. Based on Bayesian rules, when $o_{i}$ is observed, the posterior distribution of action $a_{i}$ is:
\begin{equation}
\setlength\abovedisplayskip{6pt}
\setlength\belowdisplayskip{6pt}
b_{j}(a_{i}|o_{i})=\frac{O(o_{i}|a_{i})b_{j}(a_{i})}{\sum_{a_{i}\in A}O(o_{i}|a_{i})b_{j}(a_{i})} 
\end{equation}
where $b_{j}(a_{i})$ is the prior belief of PA action $a_{i}$ at episode $j$, and $\sum_{a_{i}\in A}O(o_{i}|a_{i})b_{j}(a_{i})$ is the normalizing constant. $b_{j}(a_{i}|o_{i})$ represents the expected probability that PA takes action $a_{i}$ when observing $o_{i}$. Note that $b_{j}(a_{i}|o_{i})$ is the conditional probability based on observation $o_{i}$, which can not be used directly as a new prior belief for the next update. Then we need to use the $b_{j}(a_{i}|o_{i})$ to update the $b_{j}(a_{i})$ to $b_{j+1}(a_{i})$, which can be used in the next episode. 

Let $A_{P}$ be the action set of PA. $\vec C_{P,j-1}=(c_{1,j-1},c_{2,j-1},$ $..., 
c_{|A_{P}|,j-1})$ denotes that the action $a_{i}$ is expected to be selected $c_{i,j-1}$ times in the past $j-1$ episode, and $\vec b_{j}(a)=(b_{j}(a_{1}),b_{j}(a_{2}),...,b_{j}(a_{|A_{P}|}))$ is the prior belief of PA actions. Considering $\sum_{a_{i}\in A_{P}}b_{j}(a_{i})=1$, the distribution of $b_{j}(a_{i})$ obeys Dirichlet distribution $Dirt(\vec b_{j}(a)|\vec C_{P,j})$ \cite{b15}.

\begin{equation}
\setlength\abovedisplayskip{2pt}
\setlength\belowdisplayskip{6pt}
Dirt(\vec b_{j}(a)|\vec C_{P,j})=\frac{\tau(\sum_{i=1}^{|A_{P}|} c_{i,j})}{\prod_{i=1}^{|A_{P}|}\tau (c_{i,j})}\prod_{i=1}^{|A_{P}|} b_{j}(a_{i})^{c_{i,j}-1}
\label{eq14}
\end{equation}
where $\tau$ is the gamma function.

At episode $j$, a new observation $o_{j}$ is received, which updates  $\vec C_{P,j-1}$ to $\vec C_{P,j}$. Then the expected prior distribution of taking action $a_{i}$ is calculated by:
\begin{equation}
\setlength\abovedisplayskip{6pt}
\setlength\belowdisplayskip{3pt}
\label{23}
	\resizebox{0.85\hsize}{!}{$\begin{split}
E(b_{j}(a_{i}))&=\frac{\sum_{k=1}^{j}b_{k}(a_{i}|o_{k})}{\sum_{k=1}^{j}\sum_{i=1}^{|A_{P}|}b_{k}(a_{i}|o_{k})}\\
&=\frac{\sum_{k=1}^{j-1}b_{k}(a_{i}|o_{k})+b_{j}(a_{i}|o_{j})}{\sum_{k=1}^{j-1}\sum_{i=1}^{|A_{P}|}b_{k}(a_{i}|o_{k})+\sum_{i=1}^{|A_{P}|}b_{t}(a_{i}|o_{j})}\\
&=\frac{c_{i,j-1}+b_{j}(a_{i}|o_{j})}{\sum_{i=1}^{|A_{P}|}(c_{i,j-1}+b_{j}(a_{i}|o_{j}))}
	\end{split}$}
\end{equation}

Meanwhile, $Dirt(\vec b_{j-1}(a)|\vec C_{P,j-1})$ is the Dirichlet distribution, and the expected value of $b_{j-1}(a_{i})$ is:
\begin{equation}
\setlength\abovedisplayskip{3pt}
\setlength\belowdisplayskip{3pt}
\resizebox{0.85\hsize}{!}{$\begin{split}
E(b_{j-1}&(a_{i}))=\int b_{j-1}(a_{i})Dirt(\vec b_{j-1}(a)|\vec C_{P,j})d(\vec b_{j-1}(a))\\
&=\frac{\tau(\sum_{i=1}^{|A_{P}|} c_{i,j-1})}{\prod_{i=1}^{|A_{P}|}\tau (c_{i,j-1})}\int b_{j-1}(a_{i})^{c_{i,j-1}+1-1}\\ \prod_{k=1,k\not=i}^{|A_{P}|}&b_{j-1}(a_{k})^{c_{k,t-1}-1} d(\vec b_{j-1}(a)) \\
&=\frac{\tau (c_{i,j-1}+1)\prod_{k=1,k\not=i}^{|A_{P}|}\tau (c_{i,j-1})}{\tau(1+\sum_{i=1}^{|A_{P}|} c_{i,j-1})}
\frac{\tau(\sum_{i=1}^{|A_{P}|} c_{i,j-1})}{\prod_{i=1}^{|A_{P}|}\tau (c_{i,j-1})}\\
&=\frac{c_{i,j-1}}{\sum_{i=1}^{|A_{P}|}c_{i,j-1}}\\
\end{split}$}
\end{equation}

In Dirichlet distribution, with new experiment result $\Delta b$, the new expectation becomes:
\begin{equation}
\setlength\abovedisplayskip{3pt}
\setlength\belowdisplayskip{3pt}
\label{25}
E(b_{j}(a_{i}))=\frac{c_{i,j-1}+\Delta b}{\sum_{i=1}^{|A_{P}|}(c_{i,j-1}+\Delta b)}
\end{equation}

\begin{figure}[!t]
\setlength{\abovecaptionskip}{0pt} 
\centering
\includegraphics[width=9cm,height=2.3cm]{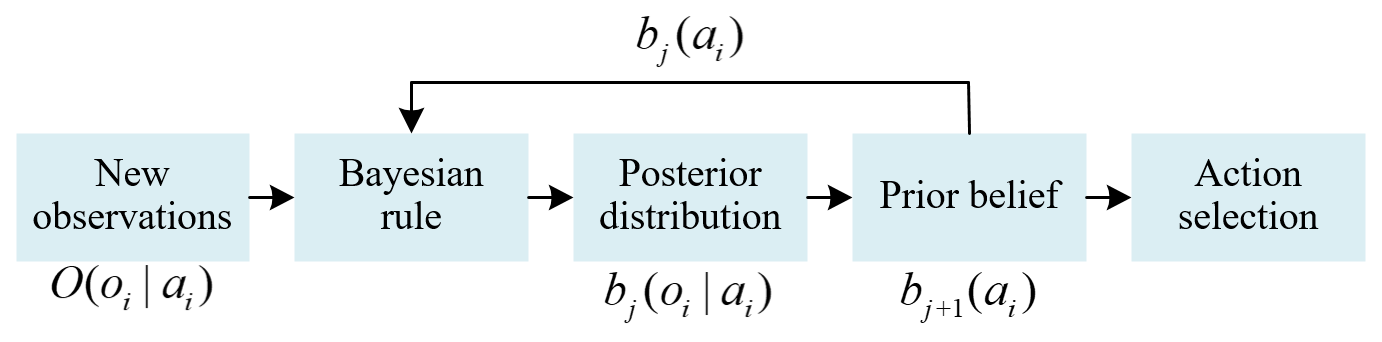}
\caption{Bayesian belief update method.}
\label{fig1-12}
\end{figure}

It is obvious that $\Delta b=b_{j}(a_{i}|o_{j})$ if we compare equation (\ref{25}) with (\ref{23}), which means we could use $b_{j}(a_{i}|o_{j})$ to update the expectation of taking action $a_{i}$, and $E(b_{j}(a_{i}))$ is used as a new prior distribution of $a_{i}$ in (\ref{25}). 

The proposed Bayesian belief update method can be summarized by Fig.\ref{fig1-12}. Given observations $O(o_{i}|a_{i})$ and prior belief $b_{j}(a_{i})$,  Bayesian rule is applied to calculate the posterior distribution $b_{j}(a_{i}|o_{i})$. Then, we update the $b_{j+1}(a_{i})$ by equation (13) to (16), which will be used in the next episode to calculate new posterior distribution $b_{j+1}(a_{i}|o_{i})$. By repeating this cycle, the prior belief on the PA actions can be used in the action selection of BA-DRL, which will be shown in Section V-B.

\label{s5-1}
\begin{figure*}[!t]
\setlength{\abovecaptionskip}{0pt} 
\centering
\includegraphics[width=17cm,height=7.7cm]{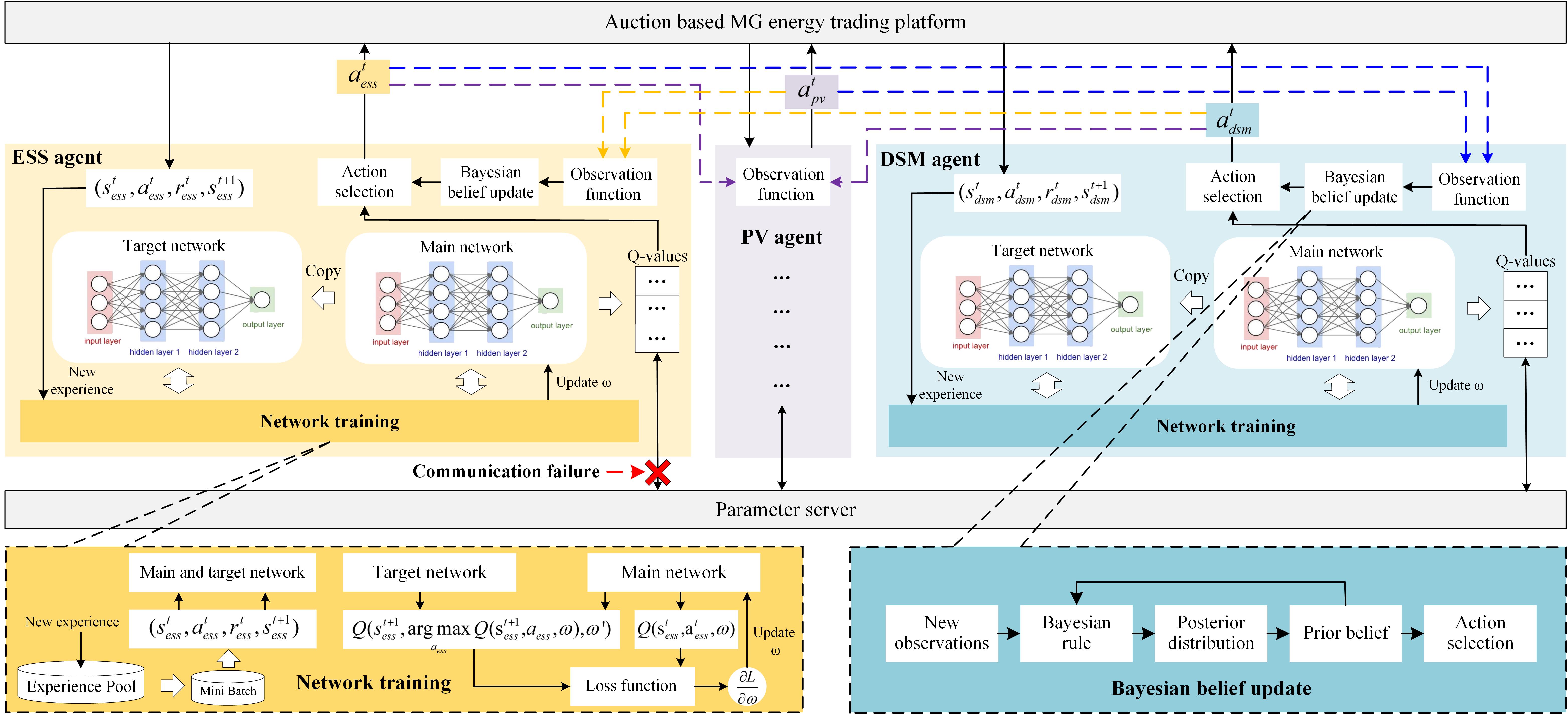}
\caption{Multi-agent BA-DRL architecture.}
\label{fig4}
\end{figure*}

\section{Bayesian Deep Reinforcement Learning}
\label{s5}
Based on the defined MA-POMDP, we present the BA-DRL algorithm. In this section, we first introduce the overall architecture and network training method of BA-DRL, and then we propose a belief-based correlated equilibrium for the action selection under communication failure. Moreover, Nash-DQN and ADMM methods are introduced as baseline algorithms.

\subsection{BA-DRL Architecture and Network Training}

In conventional Q-learning, the Q-values are updated according to: 
%\begin{small}
\begin{equation}
\setlength\abovedisplayskip{2pt}
\setlength\belowdisplayskip{2pt} \label{17-1}
\begin{aligned}
Q^{new}(s^{t},a^{t})&= Q^{old}(s^{t},a^{t})+\\&\alpha(r^{t}+\gamma \max\limits_{a} Q(s^{t+1},a) -Q^{old}(s^{t},a^{t}))
\end{aligned}
\end{equation}
%\end{small}
where $\alpha$ is the learning rate, $\gamma$ is the discount factor, $s$ is the state, $a$ is the action, $Q^{old}(s^{t},a^{t})$ and $Q^{new}(s^{t},a^{t})$ are old and new Q-values. When the Q-values converge, we have  $Q^{old}(s^{t},a^{t})=Q^{new}(s^{t},a^{t})$ in equation (\ref{17-1}), which means $Q^{old}(s^{t},a^{t})=r^{t}+\gamma \max\limits_{a} Q(s^{t+1},a)$. Then we can define a loss function for the network training of DQN:
\begin{equation}
\setlength\abovedisplayskip{5pt}
\setlength\belowdisplayskip{5pt} \label{17}
L(w)=Er(r^{t}+\gamma \max\limits_{a} Q(s^{t+1},a,w')-Q(s^{t},a^{t},w))
\end{equation}
where $w$ and $w'$ are the weight of main and target networks, respectively, $Er$ is the error function. $r^{t}+\gamma \max\limits_{a} Q(s^{t+1},a,w')$ indicates target Q-values are generated by target network, and $Q(s^{t},a^{t},w)$ means current Q-values are predicted by main network. However, in DQN, the max operator will select overestimated values, leading to overoptimistic estimation\cite{b16}. To this end, the DDQN has been proposed in which action selection and evaluation are decoupled. The loss function of DDQN is defined as:
\begin{equation}
\setlength\abovedisplayskip{3pt}
\setlength\belowdisplayskip{3pt} \label{17-2}
\begin{aligned}
L&(w)=Er(r^{t}+ \\
&\gamma Q(s^{t+1},\arg \max\limits_{a}Q(s^{t+1},a,w),w') - Q(s^{t},a^{t},w))
\end{aligned}
\end{equation}
where target network will evaluate the action, and main network will select actions by $\arg \max\limits_{a}Q(s^{t+1},a,w))$. By decoupling the action selection and evaluation, the DDQN can provide more accurate Q-values than DQN.

With the DDQN architecture, the proposed multi-agent BA-DRL is shown in Fig.\ref{fig4}, where each agent runs BA-DRL independently. For example, in ESS agent shown in yellow, after $a^{t}_{ess}$ is sent to MG trading platform, a tuple $(s^{t}_{ess},a^{t}_{ess},r^{t}_{ess},s^{t+1}_{ess})$ is received from MG trading environment, which will be stored in the experience pool. Then the network training part is shown at the bottom. ESS agent implements a random minibatch from experience pool. For every data tuple $(s^{t}_{ess},a^{t}_{ess},r^{t}_{ess},s^{t+1}_{ess})$, main network predicts $Q(s^{t}_{ess},a^{t}_{ess},w)$ and finds actions by $\arg \max\limits_{a_{ess}}Q(s^{t+1}_{ess},a_{ess},w))$. Target network evaluates the action by $Q(s^{t+1}_{ess},\arg \max\limits_{a_{ess}}Q(s^{t+1}_{ess},a_{ess},w),w')$. Predicted values are utilized in loss function by gradient descent to update main network weight. After some iterations, main network weight will be copied to the target network. This late update provides a stable reference for main network, which helps BA-DRL to be robust to power supply uncertainty such as PV power\cite{b11-6}. 

Meanwhile, here we use the Long Short Term Memory (LSTM) structure in target and main networks, which is a special recurrent neural network (RNNs). The hidden node of traditional RNNs only include a single activation function, but the hidden node in LSTM is a memory cell with forget, input, and output gates, and contents can be memorized, erased or exposed accordingly \cite{b16-1}. As such, LSTM can better capture the long-term data dependencies. Compared with existing DDQN frameworks that applied traditional deep neural networks \cite{b11-6,b16} , our BA-DRL uses LSTM to better learn the complicated long-term MG energy trading patterns. 

Finally, in Fig.\ref{fig4}, note that the action selection part is affected by both beliefs and Q-values. On one hand, ESS agent will collect observations of $a^{t}_{PV}$ and $a^{t}_{DSM}$, which is shown by the yellow dashed lines on the top. Given the observations as input, the Bayesian belief update is illustrated by the block at the bottom of Fig.\ref{fig4}, and the updated prior beliefs will be finally used for action selection. 
On the other hand, main network will predict Q-values of current state, and agents exchange Q-values by the parameter server for collaboration. But the communication failure may occur (shown by the red cross at the bottom), and ESS agent (or another agent) can be temporarily isolated. However, with the Bayesian belief update method defined in Section \ref{s4-2}, agents can use beliefs to select actions under communication failure, which will be introduced in next section.

\subsection{Belief-based Correlated Equilibrium}
\label{s5-2}
Action selection is a critical part of reinforcement learning algorithms. For one agent case, the $\epsilon$-greedy policy is generally applied for action selection. However, multi-agent action selection is much more complicated because the environment is affected by the actions of multiple agents simultaneously. With communication failures, some agents will lose collaboration information with the community, and it is hard to make a optimal joint-action for all agents. Although the collaboration information of isolated agents is missing, normal agents can still utilize their beliefs to estimate possible actions of isolated agents and make the best response accordingly.

In this section, we propose a belief-based correlated equilibrium, which enables normal agents to use beliefs to make the best response for communication failure in a decentralized manner\cite{b17}. When the agent $\lambda$ is isolated due to a communication failure, other agents choose the joint-action by:
\begin{equation}
\setlength\abovedisplayskip{2pt}
\setlength\belowdisplayskip{3pt}
\label{26}
\resizebox{0.9\hsize}{!}{$\begin{aligned}
\max_{a_{\lambda}\in A_{\lambda}} b(a_{\lambda}) \sum_{\vec a_{-\lambda} \in A_{-\lambda}} Pr(s,\vec a_{-\lambda})  Q(s,\vec a_{-\lambda},a_{\lambda})  \quad\quad \quad\\
sub. to \sum_{\vec a_{-\lambda} \in A_{-\lambda}}Pr(s,\vec a_{-\lambda})=1 \quad \quad \quad\quad \quad\quad\quad \quad\\
\sum_{a_{-\lambda,-k}\in A_{-\lambda,-k} }Pr(s,\vec a_{-\lambda})(Q(s,\vec a_{-\lambda})-Q(s,\vec a_{-\lambda, -k}, a_{\lambda},a_{k}))\geq0\\
0 \leq Pr(s,\vec a_{-\lambda})\leq 1 \quad \quad \quad \quad \quad\quad \quad \quad\quad\quad\quad \quad\\
\end{aligned}$}
\end{equation}
where $b(a_{\lambda})$ is the belief that agent $\lambda$ will choose action $a_{\lambda}$. $\vec a_{-\lambda}$ means the action combination of other $n-1$ agents except agent $\lambda$, $Pr(s,\vec a)$ is the probability of taking joint-action $\vec a=(a_{1},a_{2},...,a_{n})$ under state $s$, $A_{-\lambda}$ is the set of $\vec a_{-\lambda}$. 

The objective of (\ref{26}) is to maximize the total expected reward of NAs, and find an optimal probability distribution to enable each agent to choose an optimal action. We multiply the belief $b(a_{\lambda})$ with the total expected reward of other $n-1$ agents $\sum_{\vec a_{-\lambda} \in A_{-\lambda}} Pr(s,\vec a_{-\lambda})  Q(s,\vec a_{-\lambda},a_{\lambda})$ to represent the expected reward under this belief. Meanwhile, the constraints guarantee that action combination $\vec a_{-\lambda}$ brings more expected reward with probability distribution $Pr(s,\vec a_{-\lambda})$, which is indicated by $Pr(s,\vec a_{-\lambda})(Q(s,\vec a_{-\lambda})-Q(s,\vec a_{-\lambda, -k}, a_{\lambda},a_{k}))\geq0$. 

The BA-DRL is summarized in Algorithm 1. To realize the optimization constraints in the learning process, we apply the action selection constraints here. For example, ESS agent cannot choose to discharge if $SOC=0$.

\begin{algorithm}[!tb]
	\caption{BA-DRL}%算法标题
	\begin{algorithmic}[1]%一行一个标行号
		\STATE \textbf{Initialize:} MG and BA-DRL parameters
		\FOR{$j=1$ to $episode$}
		  \FOR{$t=1$ to $T^{op}$}
	     	\FORALL{Agents}
	          	\STATE With probability $\epsilon$, choose actions randomly; otherwise, predict $Q(s^{t},a^{t},w)$, and:
	        	\IF{Communication Failure}
	        	    \STATE  PAs select actions by greedy policy.
	            	\STATE NAs exchange Q-values and calculate belief-based correlated equilibrium by equation (\ref{26}).
	        	\ELSE
	            	 \STATE All agents exchange Q-values and find optimal joint-action by correlated equilibrium. 
	         	\ENDIF
	         	\STATE Make observations of other agents' actions. Calculate the posterior distribution and update $\vec C_{P,j}$ by equation (\ref{23}).
	    	\ENDFOR
		\STATE Agents take actions and update its own state, and save $(s^{t},a^{t},r^{t},s^{t+1})$ to their own experience pool.
		\ENDFOR
		\STATE Every $C$ episodes, random sample a minbatch from experience pool. Generate target Q-values $Q^{T}(s^{t},a^{t})$= 
		\begin{small} \begin{equation} \notag 
             \left\{
             \begin{array}{ccl}
            r^{t} \qquad \qquad &   if\;done\\
            r^{t}+ \gamma Q(s^{t+1},\arg \max\limits_{a}Q(s^{t+1},a,w),w')  &  else\\
            \end{array} \right.
            \end{equation}\end{small}
		\STATE Update $w$ using gradient descent by minimizing the loss $L(w)=Er(Q^{T}(s^{t},a^{t})-Q(s^{t},a^{t},w))$.
		\STATE Copy $w$ to $w'$ after several training.
		\ENDFOR
	\STATE \textbf{Output:}Optimal action sequence from $t=1$ to $T^{op}$	
	\end{algorithmic}
\end{algorithm}

\subsection{Baseline Algorithms: Nash-DQN and ADMM }
\label{s5-3}

\begin{algorithm}[!t]
	\caption{Nash-DQN}%算法标题
	\begin{algorithmic}[1]%一行一个标行号
		\STATE \textbf{Initialize:} MG and DQN parameters
		\FOR{$j=1$ to $episode$}
		\FOR{$t=1$ to $T^{op}$}
		\FORALL{Agents}
		\STATE With probability $\epsilon$, choose actions randomly; otherwise, predict $Q(s,a,w)$ and find its own Nash equilibrium by equation (\ref{27}).
		\ENDFOR
		\STATE Agents exchange equilibrium information and match the system equilibrium iteratively. Neglect the isolated agent if communication failure occurs.
		\STATE Agents take actions and update its own state $s$, and save $(s^{t},a^{t},r^{t},s^{t+1})$ to their own experience pool.
		\ENDFOR
		\STATE Every $C$ episodes, random sample a minbatch from experience pool, the target Q-values $Q^{T}(s^{t},a^{t})$ is: 
		\STATE \begin{small} \begin{equation} \notag 
             \left\{
             \begin{array}{ccl}
            r^{t} &   if\;done\\
            r^{t}+ \gamma \max\limits_{a} Q(s^{t+1},a,w')  &  else\\
            \end{array} \right.
            \end{equation}\end{small}
		\STATE Update $w$ using gradient descent by minimizing the loss $L(w)=Er(Q^{T}(s^{t},a^{t})-Q(s^{t},a^{t},w))$.
		\STATE Copy $w$ to $w'$ after several trainings.
		\ENDFOR
	\STATE \textbf{Output:}Optimal action sequence from $t=1$ to $T^{op}$
	\end{algorithmic}
\end{algorithm}

In this paper, we use the Nash-DQN and ADMM as baseline algorithms. Nash equilibrium is generally applied for multi-agent coordination problem, and Nash-DQN is compared as a learning-based algorithm\cite{b10}.
Meanwhile, ADMM algorithm is a well-known model-based distributed optimization method, which can be applied for MG economic dispatch problems \cite{b3}.

Nash-DQN is a multi-agent DRL method, in which each agent runs a DQN independently and Nash equilibrium is applied for joint-action selection\cite{b10}. In Nash equilibrium, no agent will change its action because it will result in a lower utility. The Nash equilibrium is described as:
\begin{equation} \label{27}
U_{\lambda}(\vec a_{-\lambda},a_{\lambda})\geq U_{\lambda}(\vec a), \; \vec a\in A,  \;\vec a_{-\lambda}\in A_{-\lambda},
\end{equation}
where $U_{\lambda}$ is the utility function for agent $\lambda$, $\vec a$ is the joint-action of all agents, $\vec a_{-\lambda}$ is the joint-action of other agents except agent $\lambda$, $A$ is the action set of $\vec a$, and $A_{-\lambda}$ is the action set of $\vec a_{-\lambda}$. Equation (\ref{27}) indicates that agent $\lambda$ will not change its action $a_{\lambda}$ because it will lead to a lower utility $U_{\lambda}$ for itself. Furthermore, the multi-agent system will reach Nash equilibrium when all agents get their own equilibrium. It is worth noting that all agents need to exchange information iteratively to find a system level Nash equilibrium, which increases the risk of communication failure. Here we assume other agents will neglect the isolated agents under communication failure. Nash-DQN is given in Algorithm 2. 

On the other hand, here we compare the ADMM with BA-DRL to show the superiority of learning-based method. In the ADMM framework, the MG system is described as:
\begin{equation}
\setlength\abovedisplayskip{3pt}
\setlength\belowdisplayskip{3pt}
\label{28}
\begin{aligned}
\min \;cost^{dsm}(P^{dsm}_{t})-profit^{pv}(p^{pv}_{t})\\
\quad\quad \quad-profit^{ess}(P^{ess}_{t},p^{ess}_{t})\\
sub. to\; (\ref{10})-(\ref{12})\quad \quad \quad\quad \quad\quad\quad \quad \\
\end{aligned}
\end{equation}
where $P^{dsm}_{t}$ and $P^{ess}_{t}$ are power of DSM and ESS agents, respectively. $p^{pv}_{t}$ and $p^{ess}_{t}$ are the biding prices of PV and ESS agent at time $t$. The convexity of equation (\ref{28}) can be easily proven as a combination of quadratic and linear problems. In ADMM, the problem will be first transformed to an augmented Lagrangian form. Then each sub-objective is updated in a sequential manner, and the iteration will stop when the value converges \cite{b18}. Noting that agents need to exchange dual variables frequently in ADMM. We assume ADMM is applied in each time slot until convergence, then the next time slot comes.

\section{Performance Evaluation}
\label{s6}
\subsection{Simulation Settings}

\begin{table}[!t]
\setlength{\abovecaptionskip}{0pt} 
\caption{Parameter Settings.}
\centering
\renewcommand\arraystretch{1.3}
\begin{tabular} {|m{2.1cm}<{\centering}|m{1.7cm}<{\centering}||m{2cm}<{\centering}|m{1.3cm}<{\centering}|} \hline
\multicolumn{2}{|c||}{MG Parameters} & \multicolumn{2}{c|}{BA-DRL Parameters} \\ \hline
PV capacity & 30kW & Network layers & 4\\ \hline
PV prediction error & $N(0,0.1P_{t}^{pv})$ & Training frequency & Every 40 episodes \\ \hline
ESS charging power & 20kW  & Hidden layers  & 2 LSTM (35 nodes)  \\ \hline
ESS capacity  & 120kW·h & Batch size & 120\\ \hline
ESS initial SOC  & 0.4  & Experience Pool & 1200\\ \hline

SOC limit & [0,1] & Learning rate & 0.005\\ \hline
Grid feed in price & 40\% & Discount factor & 0.6\\ \hline
\end{tabular}
\label{tab2} 
\vspace{0pt}
\end{table}

\begin{table}[!t]
\vspace{0pt}
\setlength{\abovecaptionskip}{0pt} 
\caption{Properties of deferrable devices}
\centering
\renewcommand\arraystretch{1.3}
\begin{tabular}{|p{1.7cm}<{\centering}|p{1.7cm}<{\centering}|p{1.75cm}<{\centering}|p{2.0cm}<{\centering}|}
\hline
Device Number & Average Power(kW) & Operation time limit (Hours) & Average duration time (Hours) \\
\hline
1 &6 &[1, 8]& 2\\
\hline
2 &15 &[7, 13]& 1\\
\hline
3& 19 &[10, 17]& 1\\
\hline
4& 10 &[15, 22]& 3\\
\hline
5& 7 &[20, 4$^{\mathrm{+24h}}$]& 1\\
\hline
\end{tabular}
\label{tab1} 
\vspace{0pt}
\end{table}

\begin{table}[!t]
\setlength{\abovecaptionskip}{0pt} 
\vspace{0pt}
\caption{Observation function parameters}
\centering
\renewcommand\arraystretch{1.1}
\begin{tabular}{|p{0.2cm}<{\centering}|p{0.6cm}<{\centering}|p{0.6cm}<{\centering}|p{0.6cm}<{\centering}|p{0.6cm}<{\centering}|p{0.6cm}<{\centering}|p{0.6cm}<{\centering}|p{0.6cm}<{\centering}|p{0.6cm}<{\centering}|}
\hline
&$o_{1}$&$o_{2}$&$o_{3}$&$o_{4}$& $o_{5}$&$o_{6}$& $o_{7}$&$o_{8}$  \\
\hline
$a_{1}$&0.6 & 0.0571&  0.0571& 0.0571 & 0.0571 &  0.0571&  0.0571& 0.0571  \\
\hline
$a_{2}$&0.0571 & 0.6&  0.0571& 0.0571 & 0.0571 &  0.0571&  0.0571& 0.0571  \\
\hline
$a_{3}$& 0.0571&  0.0571& 0.6 & 0.0571 &  0.0571&  0.0571& 0.0571& 0.0571  \\
\hline
$a_{4}$& 0.0571&  0.0571& 0.0571 & 0.6 &  0.0571&  0.0571& 0.0571 & 0.0571  \\
\hline
$a_{5}$& 0.0571&  0.0571& 0.0571 & 0.0571 &  0.6&  0.0571& 0.0571 & 0.0571  \\
\hline
$a_{6}$& 0.0571&  0.0571& 0.0571 & 0.0571 &  0.0571&  0.6& 0.0571  & 0.0571  \\
\hline
$a_{7}$& 0.0571&  0.0571& 0.0571 & 0.0571 &  0.0571&  0.0571& 0.6  & 0.0571  \\
\hline
$a_{8}$& 0.0571&  0.0571& 0.0571 & 0.0571 &  0.0571&  0.0571& 0.0571  & 0.6  \\
\hline
\end{tabular}
\label{tab3}
\vspace{0pt}
\end{table}

We assume there are PV, ESS and DSM agents in a MG. The predicted PV power is generated according to \cite{b11-3}, and the prediction error obeys a Gaussian distribution. The bidding price is discretized as [0.05, 0.09, 0.13, 0.17, 0.21, 0.25] \$/kW·h. MG and BA-DRL parameters are shown in Table \ref{tab2}. There are 5 sets of deferrable devices, where the operation time limit and average duration time are shown in Table \ref{tab1}. Here, $4^{+24h}$ means 4 am of the next day. 

We apply the well-known experience reply technique for the network training \cite{b18-1}. In particular, the former experience $<$state, action, reward, next state$>$ will be saved in the experience pool, which is considered as the generated data for network training. After every 40 episodes, a random minibatch is implemented in the experience pool, and the batch data will be used for main network training. The neural network contains 4 layers, including one input layer, one output layer, and two LSTM hidden layers with 35 nodes. Hyperparameter tuning is a well known issue for the network training, and here we set our values using the grid search method by trying different parameter combinations. The simulation is repeated for 10 runs in MATLAB platform to get averaged values with 95\% confidence interval. 

\subsection{MG Energy Trading with Communication Failure}

\begin{figure}[!t]
\vspace{0pt}
\centering
\subfigure[Energy trading with normal communication]{
\includegraphics[width=7.4cm,height=4.5cm]{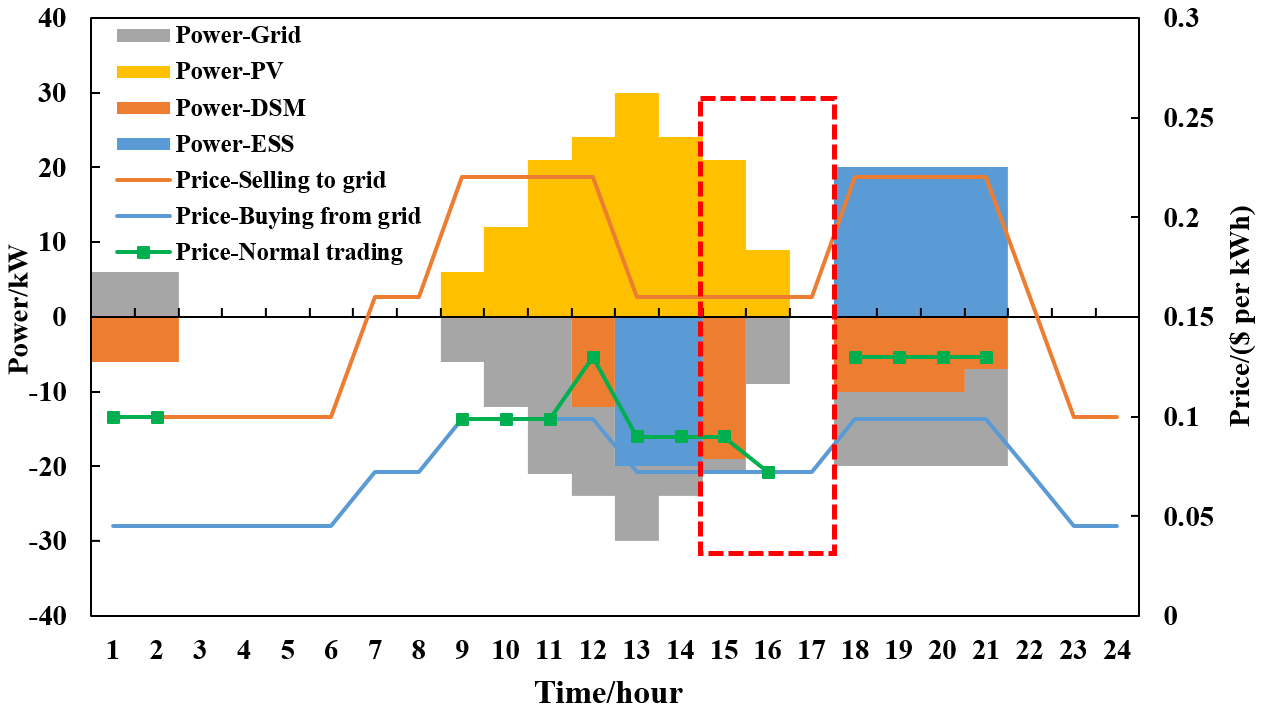}
%\caption{fig1}
}
\quad
\subfigure[Energy trading with communication failure]{
\includegraphics[width=7.4cm,height=4.5cm]{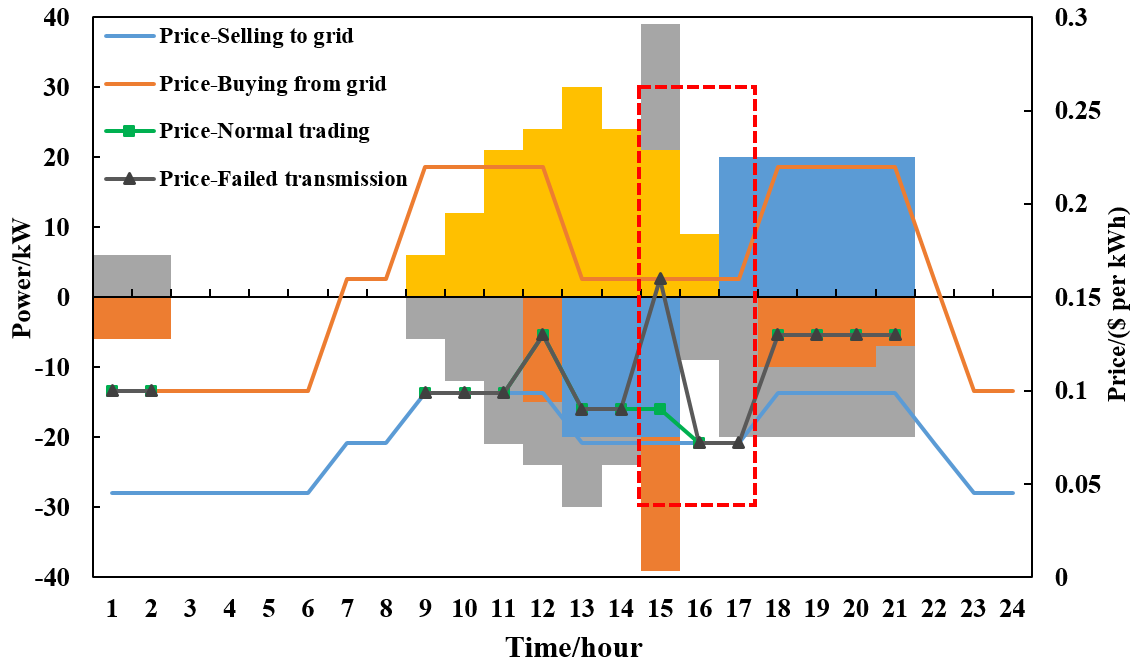}
}
\quad
\subfigure[Energy trading with Bayesian estimation]{
\includegraphics[width=7.4cm,height=4.5cm]{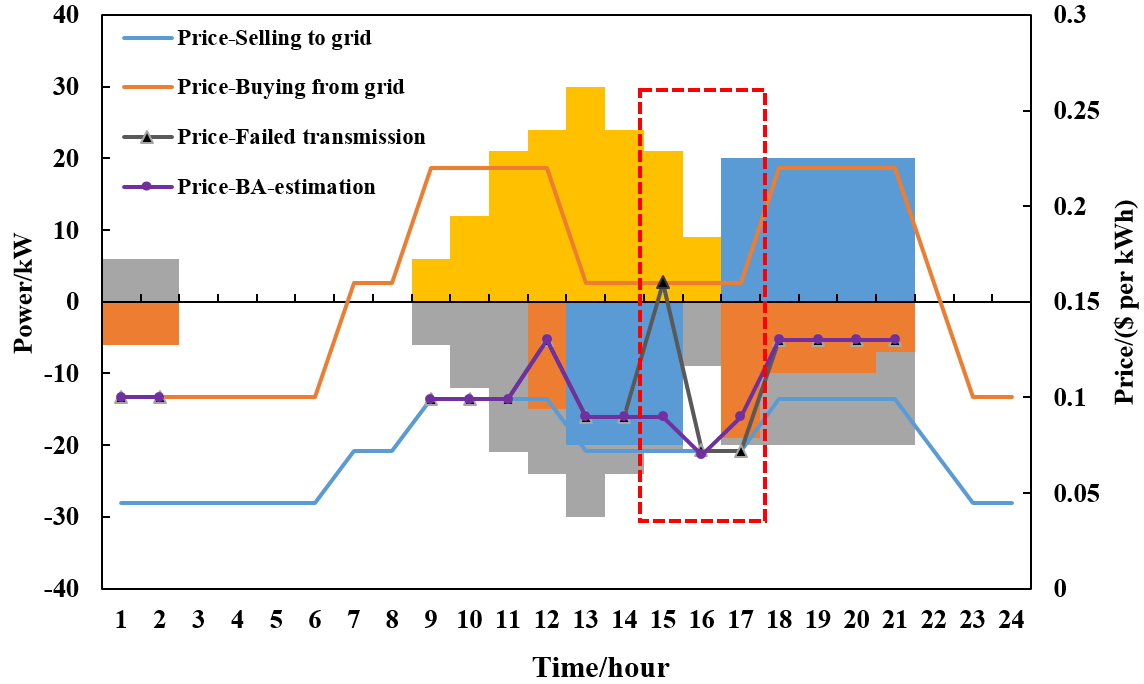}
}
\caption{MG energy management and pricing.}
\label{f8}
\end{figure}

\begin{figure*}[!t]
\centering
\subfigure[DSM cost under PV uncertainty]{
\includegraphics[width=7.4cm,height=4.5cm]{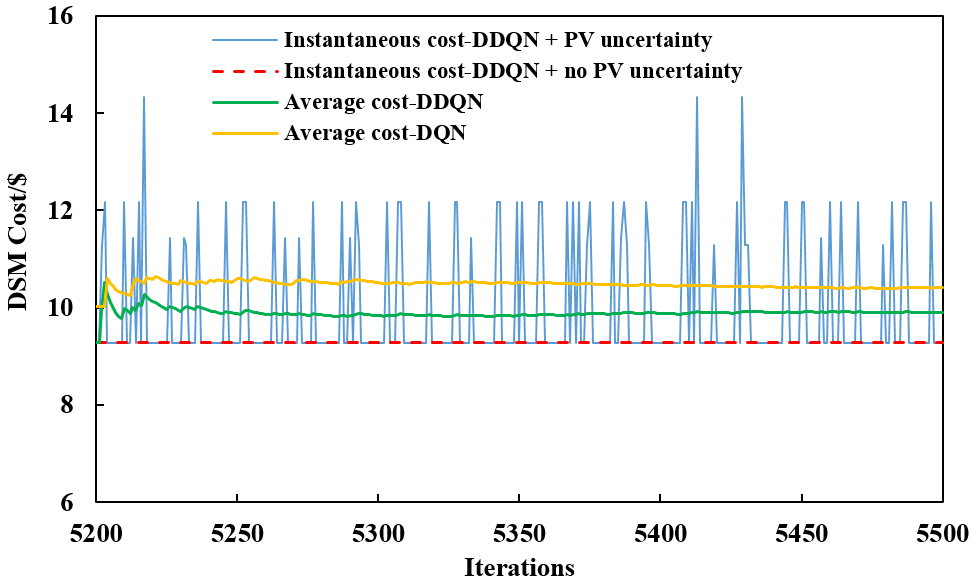}
}
\qquad
\subfigure[DSM cost under communication uncertainty]{
\includegraphics[width=7.4cm,height=4.5cm]{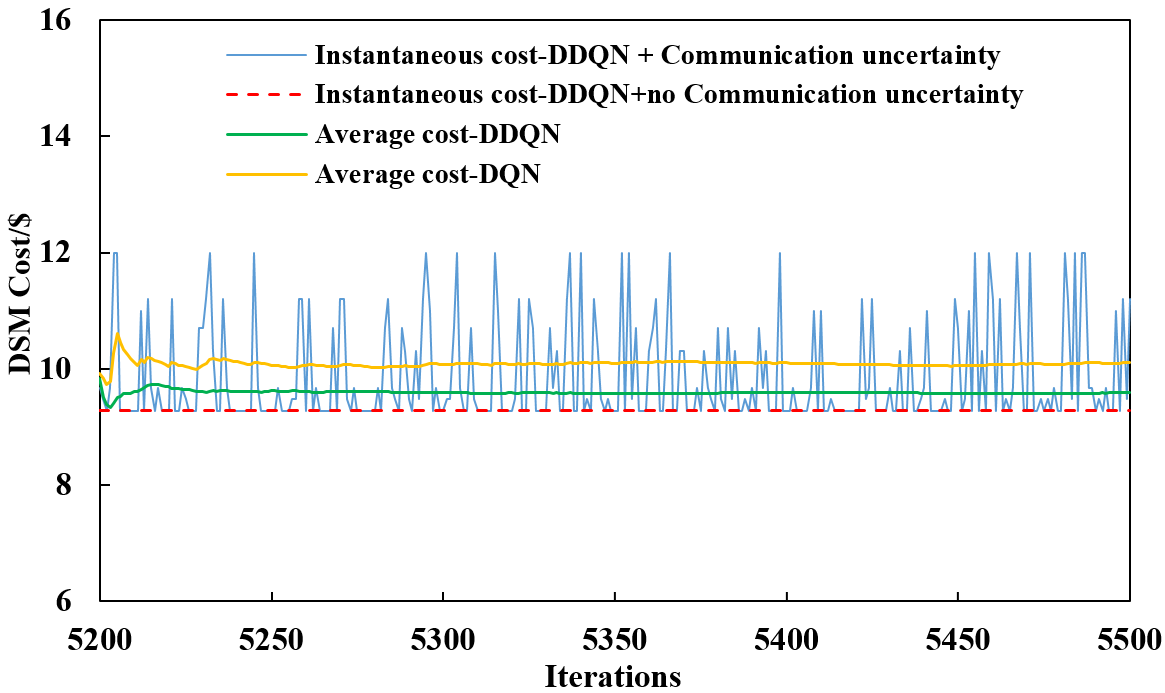}
}
\setlength{\abovecaptionskip}{0pt} 
\caption{MG system performance under uncertainty and different algorithms}
\label{f9}
\end{figure*}

In this section, we explain how MG energy trading and pricing work under BA-DRL. Fig.\ref{f8}(a) shows the performance under perfect communication. The ESS agent uses PV power to charge at daylight, and sells the electricity to DSM devices at night, which will benefit the whole MG. In the Fig.\ref{f8}(b), we assume the ESS agent suffers a communication failure at time $15$. Then ESS agent will neglect other agents and make its own decision. Compared with Fig.\ref{f8}(a), the main difference is that ESS charges at time 15. Considering that the PV power is not enough for DSM devices and ESS charging at the same time, the main grid participates in the market. Then the uniform clearing price is dominated by main grid, where the new price 0.16 \$/kW·h is much higher than the original clearing price 0.072 \$/kW·h. As a result, both ESS and DSM agents have to buy electricity at a higher price. It means that other agents are affected by the communication failure of ESS agent, and the overall profit is harmed. The failed transmission does not only change the current action, but also affects the following actions. For example, compared with Fig.\ref{f8}(a), ESS also makes a different decision at time 17 in Fig.\ref{f8}(b). 

Furthermore, we apply the BA-DRL, and PV and DSM agents will maintain beliefs on ESS's actions. Based on former definition shown as  Fig.\ref{fig1-2} in Section \ref{s4-1}, we set the observation function as Table \ref{tab3}. When action $a_{i}$ of ESS agent is actually taken, only one observation result is received. With a probability 0.6, the observation result is $o_{i}$. However, there is a chance that other observation results are received, and we assume an equal probability here. PV and DSM agents only know the observation results, but the real action is never known for them. We first explain the results under this particular observation function, and then we present the results under various observation functions in following sections. The results are shown in Fig.\ref{f8}(c). At time 15, when ESS agent makes a wrong decision, DSM agent turns off the devices, and PV agent sells electricity to ESS agent with a lower clearing price. The main reason is that DSM and PV agent make a belief that ESS agent will charge at time 15 based on their retrospective experiences, and they apply the belief-based correlated equilibrium to select joint-action. Compared with selling electricity to the grid, the ESS agent sells electricity to DSM devices at time 17, and both agents are benefited. Compared with Fig.\ref{f8}(b), the MG system buys less electricity from main grid, benefiting the MG consumers. This example helps us demonstrate how the proposed scheme works under communication failures.

\begin{figure*}[!t]
\vspace{0pt}
\centering
\subfigure[Average DSM cost comparison]{
\includegraphics[width=7.4cm,height=4.5cm]{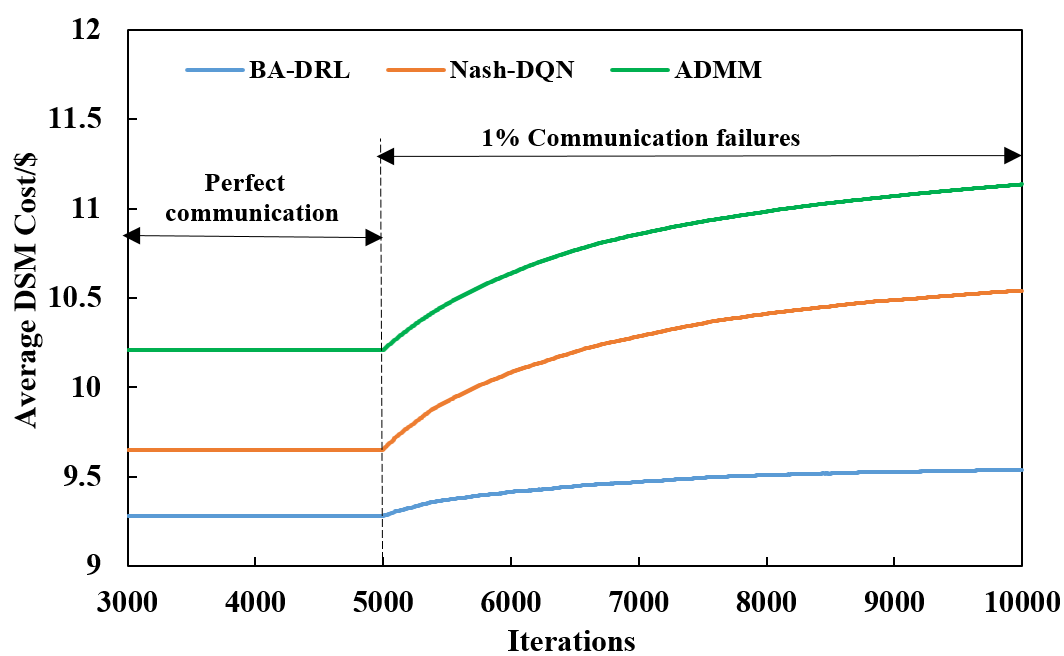}
}
\qquad
\subfigure[Instantaneous DSM cost comparison]{
\includegraphics[width=7.4cm,height=4.5cm]{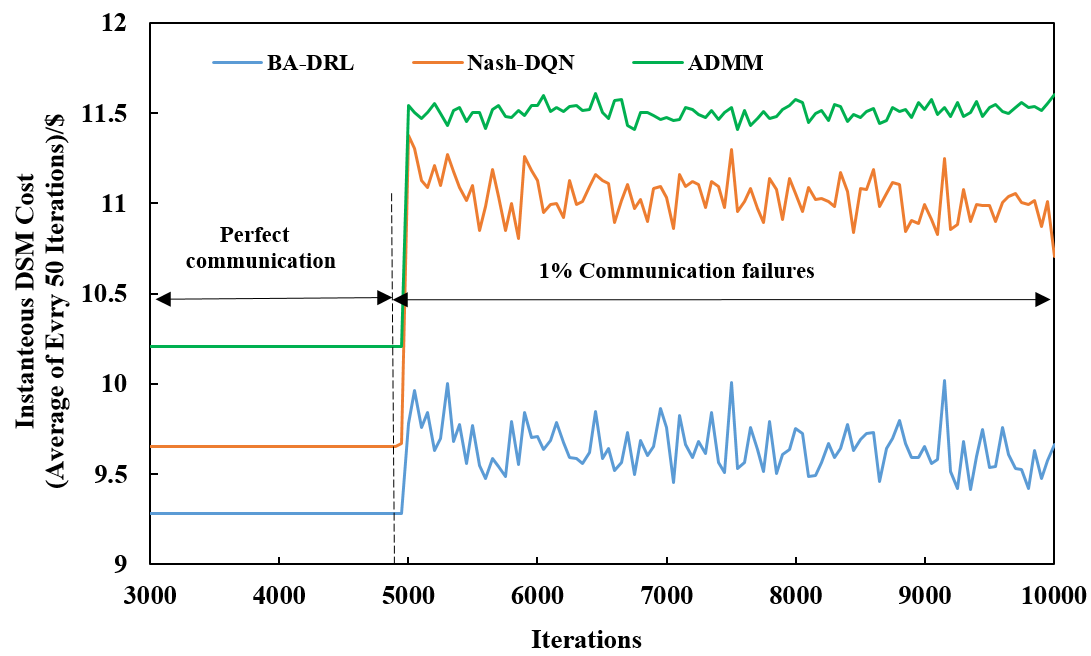}
}
\qquad
\subfigure[Average DSM cost under different observation]{
\includegraphics[width=7.4cm,height=4.5cm]{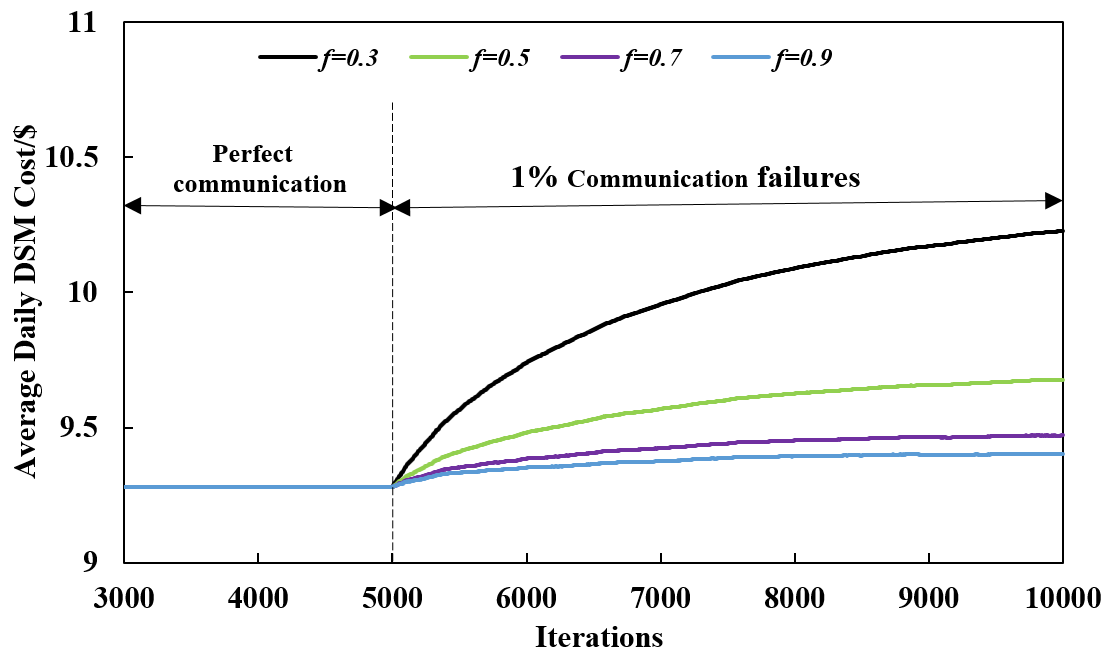}
}
\qquad
\subfigure[Estimation belief and accuracy analyses]{
\includegraphics[width=7.4cm,height=4.5cm]{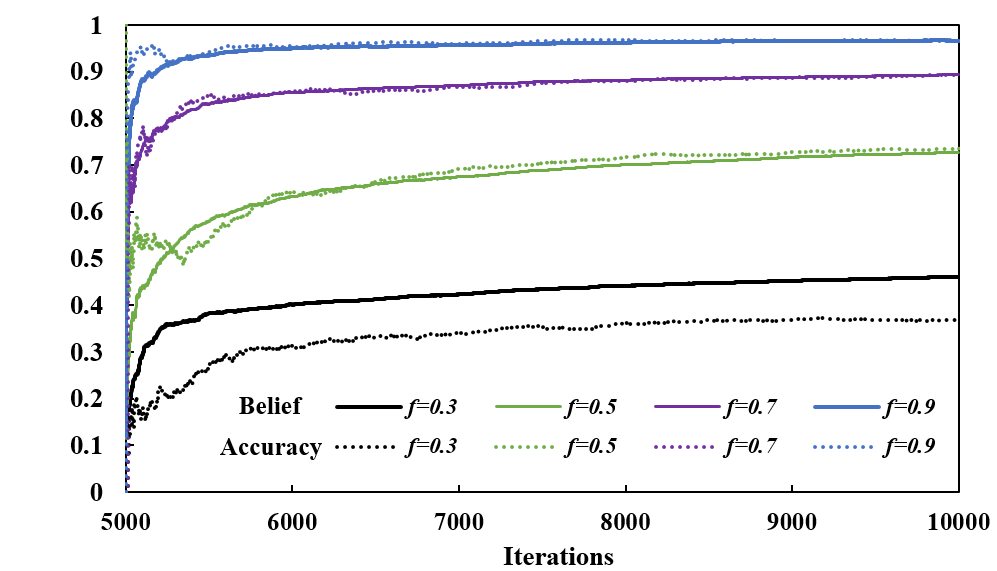}
}
\setlength{\abovecaptionskip}{0pt} 
\caption{MG system performance under uncertainty and different algorithms.}
\vspace{0pt}
\label{f10}
\end{figure*}

\subsection{BA-DQL Performance under Supply Uncertainty}

Power generation uncertainty is a critical concern for MGs that rely on renewables. Hence the designed algorithm should maintain a satisfying performance under uncertainty. In this section, we investigate the BA-DRL performance under 2 types of uncertainty: PV generation uncertainty and communication uncertainty as a result of failed transmission. Based on the BA-DRL scheme, we compare the DDQN architecture with normal DQN to present the superiority of DDQN.

The Fig.\ref{f9}(a) shows the DSM daily cost from 5200 to 5500 iterations, where each iteration contains 24 time slots as shown in Fig.\ref{f8}. The result shows that the DDQN architecture maintains a good performance under PV uncertainty. Compared with DQN architecture, the DDQN method obtains 6.7\% lower cost, and the main reason is that it overcomes the overestimation problem and provides a better estimation. Noting that the lower limit of daily DSM cost is because DSM agent has already obtained the lowest energy price. 

Fig.\ref{f9}(b) presents the performance under communication uncertainty, where there is a 1\% probability communication failure between parameter server and agents. It shows that daily DSM cost is affected by the communication failure. The missing collaboration information misleads agents, and they may make suboptimal decisions. Our proposed scheme still maintains good performance. Compared with DQN scheme, the DDQN architecture has a 5\% lower DSM cost.

The good performance of DDQN architecture in Fig.\ref{f9} (a) and (b) can be explained by its ability to mitigate the overestimation issue. In traditional DQN, both action selection and evaluation are conducted by target network, and it constantly selects the maximum Q-values of the next state. If every time the Q-value is calculated a higher value, then the Q-value predicted by neural network will be obviously higher every time, which will lead to an overoptimistic Q-value estimation. However, in DDQN, by decoupling the action evaluation and estimation, it can provide more accurate Q-values estimations, which will lead to better action selection and overall performance.

\subsection{Comparison under Communication Failure}

\begin{figure*}[!tbp]
\vspace{0pt}
\centering

\subfigure[Comparison under different algorithms.]{
\includegraphics[width=7.4cm,height=4.5cm]{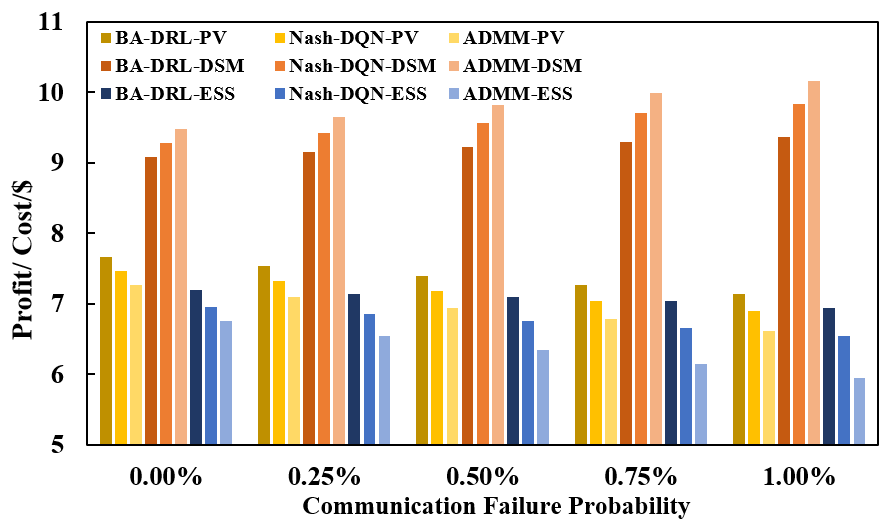}
%\caption{fig1}
}
\qquad
\subfigure[Comparison of energy exchange with grid ]{
\includegraphics[width=7.4cm,height=4.5cm]{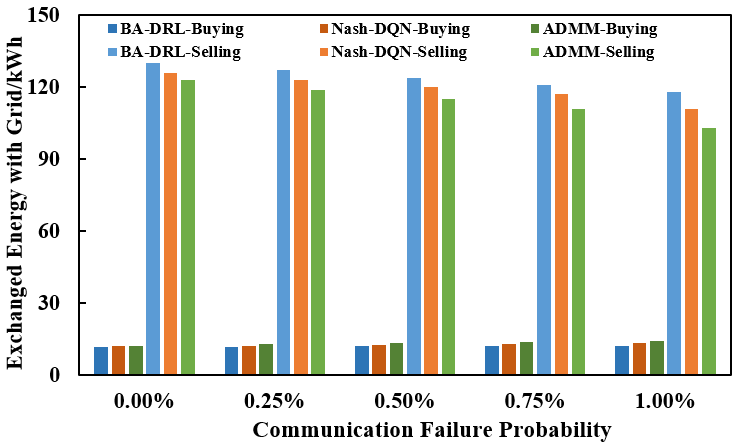}
}
\qquad
\subfigure[Comparison of MG overall reward]{
\includegraphics[width=7.4cm,height=4.5cm]{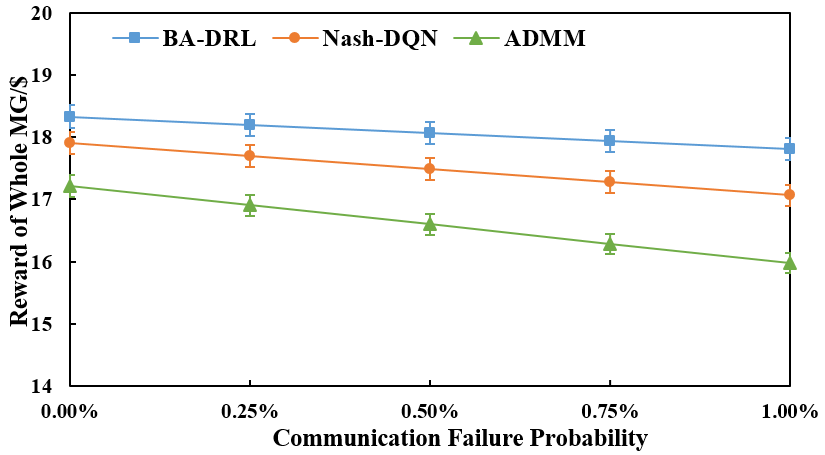}
}
\qquad
\subfigure[Comparison of number of failed transmissions]{
\includegraphics[width=7.4cm,height=4.5cm]{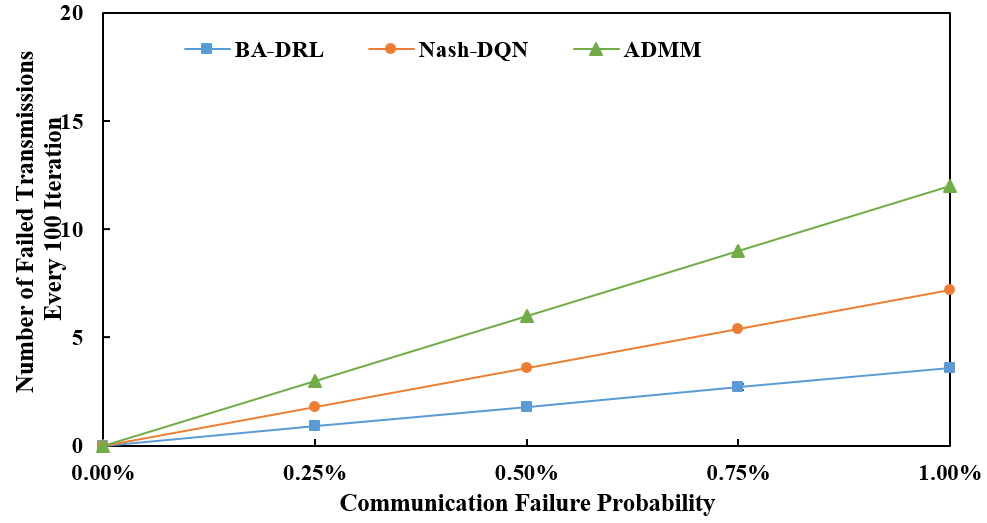}
}
\qquad
\subfigure[Comparison of different estimation strategy ]{
\includegraphics[width=7.4cm,height=4.5cm]{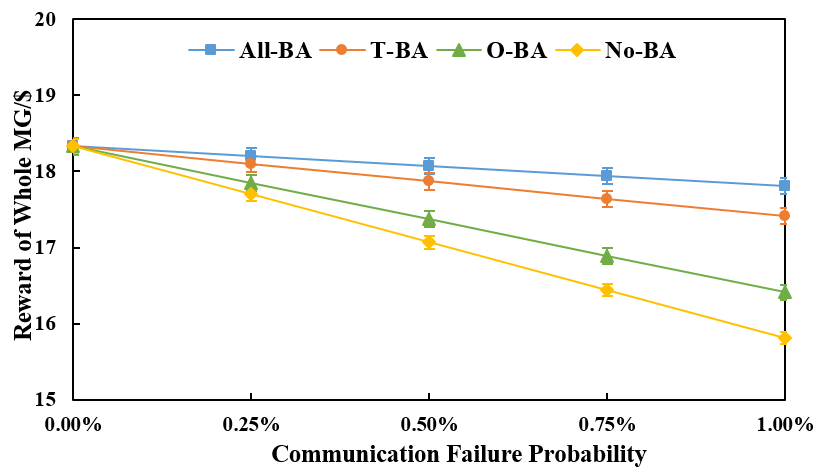}
}
\qquad
\subfigure[Comparison of different observation function]{
\includegraphics[width=7.4cm,height=4.5cm]{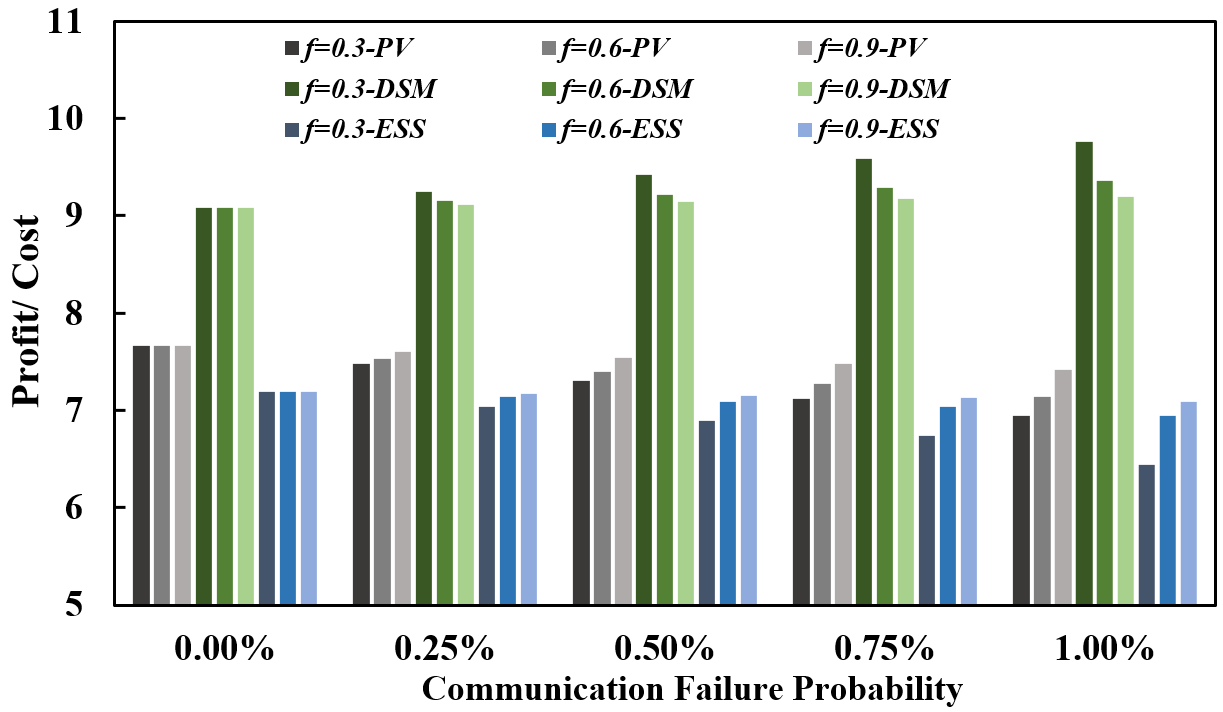}
}
\setlength{\abovecaptionskip}{2pt} 
\caption{Comparison of different communication failure probabilities}
\vspace{0pt}
\label{fig5}
\end{figure*}

In this section, we compare the BA-DRL with other baseline algorithms, including the Nash-DQN and ADMM methods. To better show the effect of communication failures, we assume normal communication between 3000-5000 episodes, and communication failure occurs with 1\% probability from 5000 to 10000 episodes.

Fig.\ref{f10}(a) and (b) show the average and instantaneous daily DSM cost. In the first 3000-5000 iterations, which is assumed to be perfect communication period, BA-DRL has a 4.3\% lower cost than Nash-DQN, and 9.1\% lower than ADMM method. The reasons are that Nash-DQN is affected by over estimation, and ADMM method is unable to detect the long-term reward. In 5000-10000 iterations, communication failures occur with 1\% probability. Then all three algorithms have a higher cost, which means failed transmissions affect the performance of DSM agent. However, the BA-DRL achieves a lower cost than Nash-DQN and ADMM. In BA-DRL, normal agents can estimate isolated agents' action pattern, and make decisions by the belief-based correlated equilibrium. On the contrary, Nash-DQN and ADMM method cannot estimate the behavior of other agents when communication fails, and the missing information leads to a higher cost.

It is obvious that the BA-DRL relies on the observation function, and Fig.\ref{f10}(c) analyzes the performance under different observation functions, where $f=0.3$ means that we use 0.3 to replace the value 0.6 in Table \ref{tab3}. Fig.\ref{f10}(d) presents the belief and accuracy, in which the accuracy means the proportion of correct estimation. Fig.\ref{f10}(c) and (d) show that a higher observation accuracy leads to lower cost and higher estimation accuracy. With more observations, agents are able to learn the action patterns of isolated agent. Based on the observation results, DSM agent updates beliefs and makes a good estimation for ESS actions under communication failures. 

\subsection{Comparison of various communication failure probabilities}

Finally, we present the performance of all agents and the whole MG under different communication failure probabilities. Fig.\ref{fig5}(a) shows that the BA-DRL achieves higher profit for PV and ESS agents, and lower cost for DSM agent. Fig.\ref{fig5}(b) shows that BA-DRL buys less energy from the grid, and sells more energy to grid, which means a higher profit for whole MG. Compared with Nash-DQN and ADMM, BA-DRL sells 5.9\% and 12.7\% more energy, and buying 8.2\% and 13.9\% less energy under 1\% communication failure probability. Fig.\ref{fig5}(c) also demonstrates that BA-DRL maintains a high reward for overall MG with varying communication failure probabilities. With 1\% communication failure probability, BA-DRL has a 4.1\% and 10.3\% higher reward compared with Nash-DQN and ADMM. 

In Fig.\ref{fig5}(d), we investigate the number of failed transmissions of different algorithms. ADMM has much more failures, and the main reason is that agents need to communicate iteratively until the result converges, while Nash-DQN has a similar problem. However, BA-DRL only exchanges Q-values once for each time slot, which causes less losses under the same failure probability. In Fig.\ref{fig5}(e), we assume the Bayesian estimation is applied in part of agents, where All-BA, T-BA, O-BA and No-BA indicate all agents, two random agents, only one agent and no agent, respectively. It demonstrates that partial estimation will lead to a lower system reward, and O-BA has a close performance to No-BA. The cost and profit of different agents under various observation function value, investigated in Fig.\ref{fig5}(f), demonstrates that an accurate observation function can lead to a higher profit and lower cost for agents.

\subsection{Convergence and Complexity analyses }

\begin{figure}[!t]
\setlength{\abovecaptionskip}{0pt} 
\centering
\includegraphics[width=7.4cm,height=5cm]{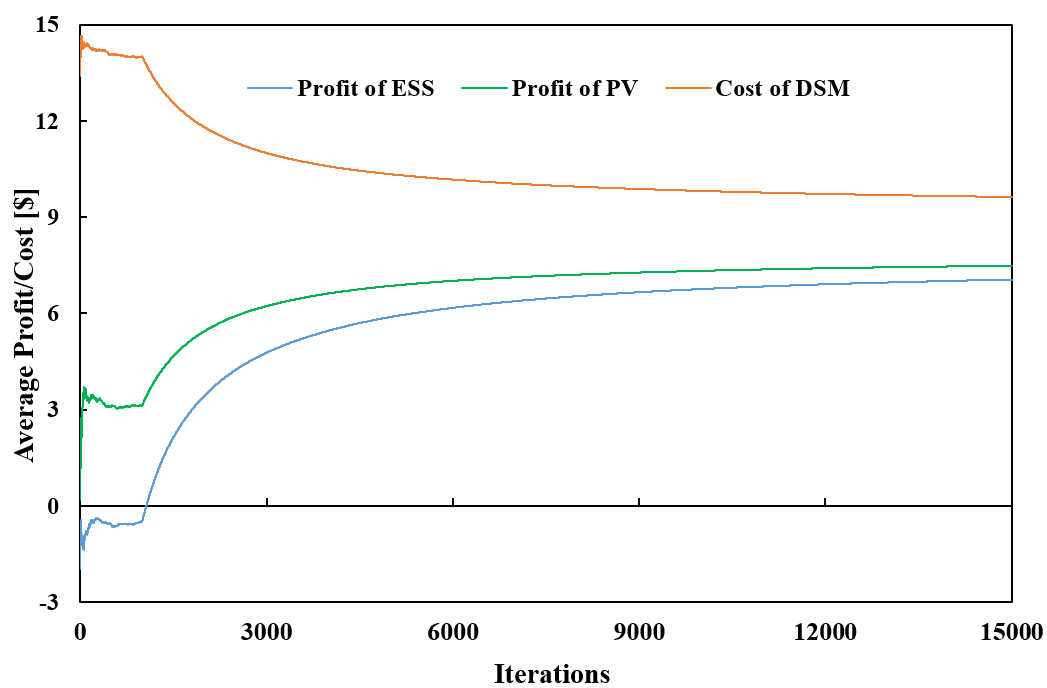}
\caption{Convergence performance of multi-agent BA-DRL.}
\label{fig23}
\end{figure}

Finally, we present the convergence performance and complexity analyses. The convergence performance of proposed multi-agent BA-DRL is shown by Fig.\ref{fig23}. It is observed that all agents benefit from the BA-DRL algorithm, including the increasing ESS and PV profits as well as the decreasing DSM cost. The results show that the proposed multi-agent BA-DRL can coordinate the action of all agents and converge to a satisfying overall performance for the whole MG. The satisfying convergence performance of BA-DRL can be explained by the good convergence property of correlated equilibrium, which has been proven in \cite{b17}. In addition, we decay the learning rate every several iterations, which contributes to a stable learning process and a good convergence performance at last.

In addition, we analyze the memory and runtime complexity of proposed algorithms. In this work, the hardware environment is Intel core i7-7770 CPU with 16 G memory size. For the memory complexity, both BA-DRL and Nash-DQN use the neural network for Q-values approximation. It means no huge Q-table is needed to record all the Q-values, and thus the memory space is greatly reduced. Meanwhile, the ADMM algorithm also has no stringent requirements for memory space, because there is no need to store variables between iterations. 

For the run time complexity, the BA-DRL is designed in a decentralized manner, and each agent runs BA-DRL parallelly, which will reduce the complexity and improve scalability. The computational complexity of LSTM network that we used in BA-DRL is given by $O((MC)^2)$ (see the appendix), where $M$ is the number of memory blocks, and $C$ is the number of memory cells in each block \cite{b18-2}. Meanwhile, the correlated equilibrium also contributes to the computational complexity. It can be easily formulated as a linear problem and solved in polynomial time $O(n^{k})$, where $n$ is the number of agents and $k$ is the power of polynomial \cite{b17,b19}. 
The average runtimes per iteration are 0.331s for BA-DRL, 0.325s for Nash-DQN and 0.129s for ADMM. The BA-DRL has a comparable runtime with Nash-DQN, but ADMM has a much lower runtime. This result can be explained by the time-consuming network training process, which is a well-known issue for DRL algorithms. However, the network training is only implemented after certain number of iterations. The runtime can be alleviated by reducing network training frequency or simplify network architecture. 

\section{Conclusion}
\label{s7}
Machine learning based and decentralized approaches are promising tools to improve energy management of MGs by allowing IoT-enabled loads, storage and generation devices act intelligently. In this paper, we proposed a BA-DRL method for MG energy management with communication failures, where a multi-agent POMDP is defined and Bayesian method is applied for action estimation and belief update. The proposed method is compared with Nash-DQN and ADMM, and a higher profit for PV and ESS agents, as well as a lower cost for DSM agent are observed under different communication failure probabilities. In the future, we will improve the scalability of proposed algorithms.  

\appendix
The computational complexity of BA-DRL is dominated by the LSTM complexity. Based on the LSTM architecture defined in \cite{b18-2}, let $M$, $C$, $I$ and $K$ represent the numbers of memory blocks, memory cells in each block, input nodes, and output nodes, respectively. The computational complexity per iteration is given by:
\begin{equation} \nonumber
\begin{split}
O(&T_{LSTM})=O(T_{r}+T_{i}+T{o})\\
&=O_{r}((M(C+2C+1))(M(C+2))+M(2C+1))\\
&+O_{i}((MC+1)K)+O_{o}((M(C+2C+1)I)\\
&=O_{r}(3M^{2}C^{2}+7M^{2}C+2M^{2}+2MC+M)\\ 
&+O_{i}(MCK+K)+O_{o}(3MCI+MI)\\
&=O_{r}(3M^{2}C^{2})+O_{i}(MCK)+O_{o}(3MCI)\\
&=O((MC)^{2}),
\end{split}
\end{equation}
where $T_{r}$ denotes the run time for recurrent connections and bias. $T_{i}$ and $T_{o}$ denote the updating time of input and output nodes, respectively.

\end{document}